**Facilitated Relaxation Overrides Energetic Pathways in a Multiphase Square-Kagome Artificial Spin Ice**

Wen-Cheng Yue[1,2,#], Zixiong Yuan[1,2,†,#], Peiyuan Huang[1,2], Ziao Wang[1,2], Hangyu Liu[3], Tianyu Li[1,2], Nuo-Zhou Yang[1,2], Jingyue Gao[1,2], Chen Ku[1,2], Shuo Song[1,2], Jiyao Shang[1,2], Yang-Yang Lyu[1,2], Yihuang Xiong[1], Xuecou Tu[1,2], Liang He[1], Xun Cao[1], Ying Dong[3,*], Huabing Wang[1,2,*], Peiheng Wu[1,2], Cristiano Nisoli[4,*], and Yong-Lei Wang[1,2,*]

[1] *Research Institute of Superconductor Electronics, State Key Laboratory of Spintronics Devices and Technologies, Nanjing University, Nanjing, China*
[2] *Purple Mountain Laboratories, Nanjing, China*
[3] *College of Metrology Measurement and Instrument, China Jiliang University, Hangzhou, China*
[4] *Theoretical Division and Center for Nonlinear Studies, Los Alamos National Laboratory, Los Alamos, NM, USA*

† Present address: Sichuan Province Engineering Research Center for Broadband Microwave Circuit High Density Integration, Chengdu, China

# Authors contributed equally

* Correspondence to: yingdong@cjlu.edu.cn; hbwang@nju.edu.cn; cristiano@lanl.gov; yongleiwang@nju.edu.cn

**Relaxation in interacting many-body systems is typically assumed to follow energy minimization, yet local dynamical rules can determine which microscopic pathways are accessible. In facilitated relaxation, these pathways are enabled or blocked by neighboring configurations. However, isolating such rules in natural materials is difficult because energetics, disorder and dynamics are usually entangled. Here we realize a geometrically tunable square-kagome artificial spin ice—a lithographic nanomagnet array—that provides a clean, designable platform for facilitated relaxation. The array hosts two tunable nearest-neighbor interactions and supports multiple low-energy phases: spin crystal, spin liquid and spin foam. In the spin-foam phase, abundant facilitator configurations allow relaxation to follow Boltzmann statistics. In the spin-crystal phase, by contrast, domain-wall excitations are trapped in non-Boltzmann distributions, although energetically preferred relaxation**

**channels exist. The ordered crystal background lacks the facilitators needed to connect topologically distinct domain-wall classes. These classes are therefore kinetically isolated, as observed in both experiment and Monte Carlo simulation. Our results show that transition-network topology can override energetic relaxation, establishing artificial spin ice as a programmable platform for constrained dynamics and nonergodic relaxation.**

1. Relaxation in interacting many-body systems is shaped not only by energetics but also by the local dynamical rules that determine which kinetic pathways are accessible [1]. In systems such as structural glasses [2], granular media [3], viral capsid assembly [4], and protein folding [5], the relaxation of a local degree of freedom often depends on the states of its neighbors, leading to cooperative dynamics [6,7], heterogeneous relaxation [8], and possibly even ergodicity breaking [1]. This conditional dynamics, known as facilitated relaxation, is central to kinetically constrained models of nonequilibrium relaxation. However, directly isolating its microscopic origin in a physical system remains challenging, because natural materials often entangle kinetic rules with disorder and complex energy landscapes.

2. A direct experimental route to such constrained dynamics requires a system in which both the elementary degrees of freedom and the local rules connecting them can be designed, imaged and statistically analyzed. Artificial spin ices (ASIs) provide such a platform. They consist of lithographically patterned arrays of magnetic nanoislands whose individual magnetization is binary owing to shape anisotropy and can therefore be described as a classical Ising spin [9–18]. These systems have enabled real-space and real-time views of phenomena emerging from geometric frustration both in and out of equilibrium [19–21], including emergent magnetic monopoles [22–25], Coulomb phases [24,26], phase transitions [27,28], thermal relaxation dynamics [29–32], and kinetically constrained behavior [29,32]. While the long-range dipolar interactions in many ASI geometries yield richly intricate energy landscapes, this inherent complexity deeply entangles energetics and dynamics, presenting a challenge when seeking to isolate kinetic constraints or directly probe facilitation.

3. To overcome this entanglement and cleanly isolate local kinetic rules, here we realize a previously unexplored ASI (Fig. 1a) patterned on a square–kagome lattice (Fig. S1). This geometry has been extensively studied in frustrated magnetism [33] and recently identified in chemically complex bulk materials [34], but has not yet been realized in a lithographic nanomagnet system. It naturally hosts two inequivalent nearest-neighbor interactions whose relative strengths can be tuned by geometry. By elongating the nanomagnets to suppress long-range dipolar interactions, we create a nearest-neighbor–dominated, Ising-like metamaterial whose energy hierarchy can be precisely controlled. This tunable system reveals a rich phase diagram and also allows for direct access to local relaxation pathways and kinetic constraints that are hard to characterize in natural materials. In particular, as we show below, in square–kagome ASI, facilitated relaxation can dominate even in the presence of well-defined energetic pathways, establishing a physical setting in which constrained dynamics, cooperative relaxation rules, and nonergodic behavior can be directly visualized and analyzed.

4. We designed and realized an artificial spin ice (Fig. 1a) composed of nanobar magnets arranged on the vertices of a square–kagome lattice (Fig. S1). As shown in Fig. 1a, the structure hosts two inequivalent nearest-neighbor interactions, $J_{\alpha\alpha}$ and $J_{\alpha\beta}$, between α–α and α–β magnet pairs, respectively, whose relative strengths are controlled by the lattice constants $l_\alpha$ and $l_\beta$. The unit cell, consisting of four three-leg vertices and with eight nanomagnets, admits 82 low-energy plaquette configurations that satisfy the quasi-ice rule (‘two-in, one-out’ or ‘two-out, one-in’, Fig. 1b) for all four vertices. These configurations can be grouped into ten topologically distinct classes (Fig. 1c and Fig. S2), where members of each class are related by rotations, mirror operations, or global spin reversal (Fig. S3). Based on their interaction energetics $E_A = 4J_{\alpha\alpha}$, $E_B = 2J_{\alpha\alpha} + 2J_{\alpha\beta}$, $E_C = 4J_{\alpha\beta}$ (the energy is set to zero for a nanomagnet pair with a head-to-tail configuration), the ten classes are further categorized into three plaquette types—A, B and C (Fig. 1c). This large, structured spin configuration space forms the basis for emergent ordering and nontrivial relaxation dynamics under different interaction hierarchies.

5. Monte Carlo simulations of the nearest-neighbor Ising model reveal three distinct low-temperature phases defined by the ratio $J_{\alpha\alpha}/J_{\alpha\beta}$: spin crystal, spin liquid, and spin foam (Fig. 1d). When α-β interactions dominate ($J_{\alpha\alpha} < J_{\alpha\beta}$), Type A plaquettes have lowest energy and a twofold degenerate, long-range ordered crystal phase composed of type-A plaquettes of alternating orientation is the ground state (Fig. S4a), leading to a spontaneous symmetry breaking transition. The corresponding magnetic structure factor map exhibits sharp Bragg peaks (Fig. 1e), confirming the formation of a spin-crystal state. When the two couplings are equal ($J_{\alpha\alpha} = J_{\alpha\beta}$), the system is maximally degenerate: all 82 ice-rule configurations, corresponding to Type A, B, and C in Fig. 1c, have the same energy and appear in proportion of their multiplicity. The system displays constrained disorder (Fig. S4b) and dynamically fluctuating behavior (Video 1), exhibiting a diffuse magnetic structure factor characteristic of a classical spin liquid (Fig. 1f). When α–α interactions dominate ($J_{\alpha\alpha} > J_{\alpha\beta}$), type-C plaquettes—with an overall multiplicity of 32—form the ground-state manifold. In this phase, $\alpha$ spins organize into local chiral arrangements that generate toroidal moments (Fig. 1c), which freeze into a highly disordered toroidal glass (Fig. S4c). In contrast, the $\beta$ spins, each situated between two frozen $\alpha$-spin toroidal moments, remain dynamically free, forming an uncorrelated gas-like subsystem (Video 2). The resulting hybrid state constitutes a spin 'foam'. The magnetic structure factor map shows highly broadened features arising from short-range $\alpha$-$\alpha$ spin correlations (Fig. 1g) in the toroidal moments.

6. The three geometry-tuned phases provide a direct route for experimentally accessing not only distinct energetic landscapes but also kinetically constrained regimes. Previous studies have demonstrated that geometric tuning can control competing interactions and collective ordering in artificial spin ice [13,27,35]. In our square–kagome ASI, tuning the relative strengths of two inequivalent nearest-neighbor interactions not only selects distinct magnetic phases but also provides access to contrasting relaxation behaviors governed by phase-dependent kinetic constraints. To experimentally validate these phases, we fabricated a series of square–kagome ASI arrays with systematically varied coupling ratios $J_{\alpha\alpha}/J_{\alpha\beta}$ by adjusting the two lattice constants $l_\alpha$ and $l_\beta$, while maintaining an approximately constant sum of the nearest-neighbor coupling strengths, $J_{\alpha\alpha} + J_{\alpha\beta}$ (Figs. S5a and S5b). This design enables us to compare the effects of relative coupling strengths on magnetic ordering and relaxation dynamics while maintaining a comparable overall interaction-energy scale. All ASI arrays were fabricated on the same silicon substrate and demagnetized simultaneously using a 72-hour AC demagnetization protocol [24,36]. Each nanomagnet was 600 nm long, 80 nm wide, and 15 nm thick. Further details of sample fabrication, the demagnetization procedure and the experimental methods are provided in the Supplemental Material. By employing relatively long nanobars with a length of 600 nm, we suppressed next-nearest-neighbor dipolar interactions to below 8% of the nearest-neighbor couplings (Fig. S6), thereby realizing a nearest-neighbor-dominated, Ising-like metamaterial that retains weak but detectable long-range interactions. Scanning electron microscopy images of representative samples are shown in Figs. 2a, 3a, and Figs. S5c-5k. Magnetic force microscopy (MFM) images acquired after a demagnetization procedure are presented in Figs. 2b, 3b, and Fig. S7, from which the spin configurations were extracted as a function of $J_{\alpha\alpha}/J_{\alpha\beta}$ (Fig. S8). Statistical analysis of these configurations reveals a continuous evolution of plaquette populations with interaction hierarchy (Figs. S9a and S9b): type-A plaquettes dominate for $\frac{J_{\alpha\alpha}}{J_{\alpha\beta}} < 1$, consistent with the spin-crystal phase; type-C plaquettes dominate for $\frac{J_{\alpha\alpha}}{J_{\alpha\beta}} > 1$, characteristic of the spin-foam phase; and all three plaquette types (A, B, and C) populate according to their degeneracies at $\frac{J_{\alpha\alpha}}{J_{\alpha\beta}} = 1$, indicating the spin-liquid regime. The corresponding experimental magnetic structure-factor maps (insets of Figs. 2c and 3c, and Fig. S10d) agree well with theoretical predictions from the Ising model (Figs. 1e-1g).

7. Although the system is designed to be dominated by nearest-neighbor interactions, its large degenerate configuration space allows weak long-range interactions to manifest under appropriate conditions. In the foam phase ($\frac{J_{\alpha\alpha}}{J_{\alpha\beta}} > 1$), nearly all the plaquettes occupy type-C configurations (Fig. 2c, Figs. S8h and S8i), which consist of four topologically distinct subclasses ($C_1$-$C_4$) with degeneracies of 4, 8, 16, and 4, respectively for a total of 32 topologies. The statistics of the plaquette populations of these subclasses from Monte Carlo simulations of the nearest-neighbor Ising Model (Fig. S9c) faithfully track the experimental data (Fig. S9d), confirming the Ising-like behavior of our system.

8. However, when normalizing the data by the plaquettes' multiplicity, differences emerge. The nearest-neighbor Ising model at low temperatures shows that populations for all four subclasses ($C_1$-$C_4$) closely track their degeneracy (Fig. 2d). This is expected, since the $\beta$ spins are completely uncorrelated, the ground state is effectively a gas of plaquettes governed by entropy-dominated statistics among degenerate configurations. In contrast, the experimental data shows systematically decreasing normalized populations from $C_1$ to $C_4$ (Fig. 2e). This deviation arises from weak but finite long-range dipolar interactions (between $\beta$ spins) that lift the degeneracy of the $C_1$-$C_4$ subclasses (as shown by the red curve of the plaquette energies in Fig. 2e), producing a small energetic hierarchy. As a result of energetic relaxation, the system settles into a population distribution consistent with Boltzmann statistics (higher energy resulting in lower population) (Fig. 2f).

9. While AC demagnetization is an out-of-equilibrium, driven kinetic process, extensive prior work has demonstrated that it effectively thermalizes artificial spin systems, yielding quasi-equilibrium distributions [24,26,37,38]. Importantly, our results in the foam phase confirm this behavior, demonstrating that the demagnetization protocol faithfully follows energetic relaxation pathways. It accurately captures even minute energy differences in the foam phase, where a gas of uncorrelated $\beta$ spins is kinetically unconstrained. This effective energetic relaxation provides a powerful benchmark and a crucial reference for the kinetically constrained, non-Boltzmann statistics observed in the crystal phase described below.

**10.** In the spin-crystal phase, the twofold-degenerate type-A ground states form ordered domains separated by domain walls composed of type-B plaquette excitations (Fig. 3c). These excitations fall into five topologically distinct classes ($B_1$-$B_5$) with degeneracies of 8, 16, 8, 8, and 8, respectively (Fig. 1c). The raw statistical populations of types $B_1$-$B_5$ excitations obtained from Monte Carlo simulations (Fig. S9e) and those extracted from experiments (Fig. S9f) may initially appear to differ. However, their degeneracy-normalized populations exhibit excellent agreement (Figs. 3d and 3e). In both cases, type-$B_1$ plaquettes are strongly suppressed, while types $B_4$ and $B_5$ dominate with nearly equal normalized populations, and types $B_2$ and $B_3$ show intermediate and equal occupations.

11. These statistics for the type-B excitations in the spin-crystal phase (Figs. 3d-3f) stand in stark

contrast to those of the type-C configurations in the spin-foam phase (Figs. 2d-2f). First, Monte Carlo simulations (Fig. 3d) based solely on the nearest-neighbor Ising model—where all five B-type excitations are exactly degenerate—do not exhibit the identical, entropy-dominated normalized population seen in the spin-foam phase (Fig. 2d). This suggests that the statistics in the spin-crystal phase are not governed by entropy. Second, as demonstrated experimentally in Fig. 2e for the foam state, weak long-range dipolar coupling plays a detectable role in our system. However, although including long-range dipolar interactions lifts the degeneracies of the B-type excitations (red curve in Fig. 3e), there is no correlation between the plaquette energies and the resulting normalized populations (Figs. 3e and 3f). This indicates that the observed statistics are also not governed by energetic relaxation. The simultaneous breakdown of both entropy- and energy-governed statistics signals a failure of Boltzmann statistics in the spin-crystal phase (Fig. 3f), pointing instead to another mechanism—kinetic constraints—as the dominant driving force.

12. To identify the microscopic origin of these kinetic constraints, we systematically analyzed the allowed transitions between the topologically distinct plaquette configurations. All transitions fall into two categories (Figs. 4a and 4b). *Direct relaxation* (Fig. 4a), induced by flipping one of the four $\alpha$-spins within a plaquette, affects only the local plaquette energy and is independent of neighboring configurations. In contrast, *facilitated relaxation* (Fig. 4b), induced by flipping a $\beta$-spin that connects two neighboring plaquettes, depends critically on the configurations of the neighboring plaquettes. Such transitions are permitted only when the neighboring plaquettes act as specific ‘facilitators’ that preserve the quasi-ice rule. Types $C_1$-$C_4$ plaquettes, whose $\alpha$ spins form circulating toroidal configurations, act as universal facilitators for all other plaquette types. In contrast, type-A plaquettes, in which $\alpha$-$\alpha$ spin pairs occupy high energy configurations, act as universal non-facilitators. Consequently, the relaxation involving $\beta$-spin flips is strongly suppressed within the ordered crystal background dominated by type-A plaquettes.

13. Mapping the low-energy transition network (Fig. 4c) reveals that the five degenerate B-type excitations separate into three kinetically disconnected groups: {$B_1$}, {$B_2$, $B_3$}, and {$B_4$, $B_5$}. Relaxation within each group proceeds via direct relaxation (α-spin flips) (Fig. 4a) and thus obeys entropy-dominated statistics, yielding nearly equal normalized populations within each group. In contrast, relaxation between groups requires facilitated pathways (Fig. 4b) that are strongly suppressed in the crystal environment full of type-A non-facilitators. As a result, $B_1$ excitations—which can relax directly into the ground state type-A configurations (right panel of Fig. 4a)—are strongly depleted, while $B_4$ and $B_5$ excitations—which require multiple facilitated steps to relax into the ground state (Fig. 4c)—exhibit the highest populations (Figs. 3d-3f).

In the foam phase, by contrast, all ground-state configurations are type-C plaquettes, which universally facilitate $\beta$-spin flips (left panel of Fig. 4b). As a result, facilitated pathways are fully accessible (Fig. 4d), and relaxation proceeds through energetically and entropically governed statistics. Given that the energy splittings between the $B_1$-$B_5$ excitations in the crystal phase (red curve in Fig. 3e) are comparable to those between the $C_1$-$C_4$ configurations in the foam phase (red curve in Fig. 2e), the stark contrast between Boltzmann statistics in the foam and non-Boltzmann statistics in the crystal demonstrates that facilitated relaxation overrides energetic pathways in the crystal phase.

14. The above results highlight a profound conceptual point: the state reached by a many-body system is determined not only by its energetic hierarchy, but also by the accessibility of the pathways connecting its configurations. In square-kagome ASI, this kinetic accessibility is encoded jointly by the lattice architecture and the interaction-selected collective phase. The two inequivalent nearest-neighbor couplings stabilize collective backgrounds with distinct facilitator populations, while the plaquette structure hosts excitations whose relaxation requires facilitated transitions. This separation of energetic preference from kinetic accessibility provides a microscopic realization of facilitated dynamics in an ordered magnetic metamaterial, illustrating how materials design can expose nonequilibrium rules that are difficult to isolate in natural systems.

15. Although facilitated relaxation has been extensively modeled in theoretical models—most notably in kinetically constrained models developed for glassy dynamics [1,39]—direct experimental access to such microscopic processes has been limited. In contrast to amorphous glass formers [40], our square-kagome metamaterial is spatially ordered and structurally well defined, allowing kinetic constraints to be distinguished from disorder and complex energetic landscapes. The resulting non-Boltzmann populations demonstrate that local dynamical constraints alone can prevent energetically favorable relaxation and kinetically isolate configurations even in the absence of quenched structural disorder.

16. In summary, we have realized a square-kagome artificial spin ice in which interaction-selected collective phases give rise to sharply contrasting relaxation dynamics. By tuning two local interactions within a fixed lattice topology, we stabilize phases with distinct facilitator populations that either enable or block energy-lowering relaxation pathways. In the spin-foam phase, abundant facilitators make these pathways widely accessible, allowing even weak energetic differences to generate Boltzmann-distributed populations. In the spin-crystal phase, by contrast, the ordered type-A background lacks the facilitators required to connect distinct domain-wall classes. Consequently, energetically favorable transitions become kinetically inaccessible, producing non-Boltzmann populations and kinetically isolating excitations without quenched disorder.

17. This artificial spin platform offers opportunities to explore facilitated dynamics beyond static final states. Time-resolved magnetic imaging, such as x-ray magnetic circular dichroism microscopy [28–30], could directly visualize cooperative relaxation events, facilitator-mediated spin flips, and dynamical heterogeneity, providing access to facilitated dynamics beyond the static final states examined here. Combined with geometry-controlled interaction hierarchies, such measurements would enable artificial spin systems to serve as programmable model systems for constrained dynamics, in which both the energy landscape and the topology of allowed kinetic pathways can be rationally engineered. Beyond magnetic systems, the geometric and kinetic principles identified here could guide investigations of constrained relaxation in superconducting vortex matter [41–44], programmable qubit arrays [45], colloidal ice [46], and mechanical metamaterials [47], where collective behavior can likewise be shaped by local rules and connectivity. These principles establish a route towards designing metamaterials in which not only the accessible states, but also the relaxation pathways connecting them, can be rationally controlled. Such pathways can be engineered to transform initialized states into distinct final output states, their controlled accessibility may also provide opportunities for physics-based computation [48].

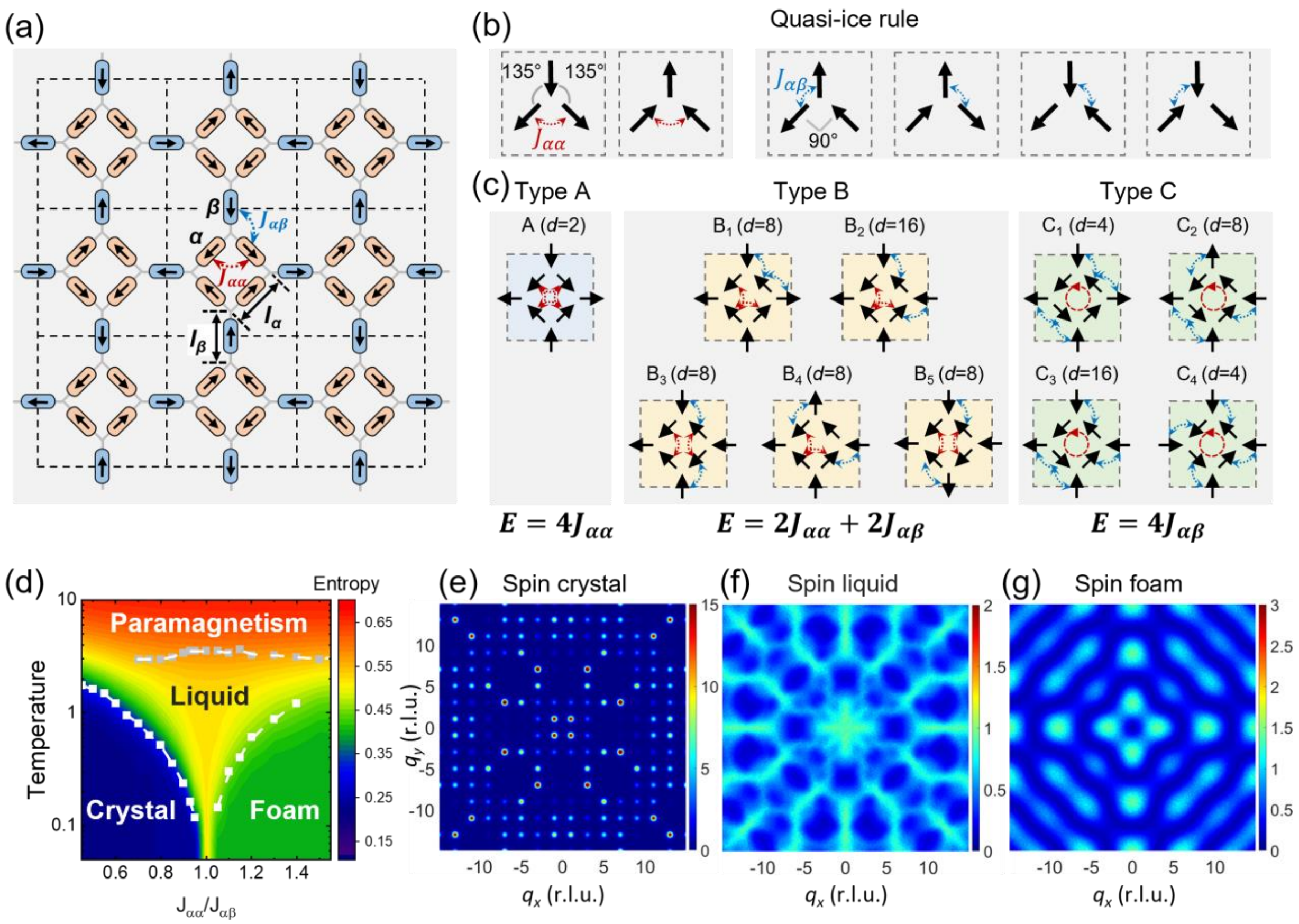

FIG. 1. Square-kagome artificial spin ice and multiple collective phases. (a) Schematic of the square-kagome artificial spin ice consisting of two interpenetrating sublattices: circulating square plaquettes formed by four α nanomagnets (orange) and a square-ice backbone formed by β nanomagnets (blue), each located between adjacent plaquettes. Nearest-neighbor couplings between α-α and α-β nanomagnets are denoted by $J_{\alpha\alpha}$ and $J_{\alpha\beta}$, respectively, with the interaction energy of a head-to-tail configuration taken as zero. The two lattice constants are labeled $l_\alpha$ and $l_\beta$. (b) Six low-energy vertex configurations that satisfy the quasi-ice rule ("two-in, one-out" or "two-out, one-in"), grouped according to the dominant interaction involving α-α and α-β nanomagnets. (c) Ten topologically distinct low-energy plaquette configurations satisfying the quasi-ice rule, classified into three plaquette types (A, B, and C) according to their interaction energies. The degeneracy $d$ of each plaquette type is indicated. (d) Phase diagram obtained from Monte Carlo simulations of the nearest-neighbor Ising model, shown as an entropy map as functions of the coupling ratio $J_{\alpha\alpha}/J_{\alpha\beta}$ and temperature. Phase boundaries (white curves) are determined from peaks in the simulated specific heat (see Fig. S12). (e)-(g) Calculated magnetic structure-factor maps for the three low-energy phases: spin crystal (e), spin liquid (f), and spin foam (g) obtained from Monte Carlo data.

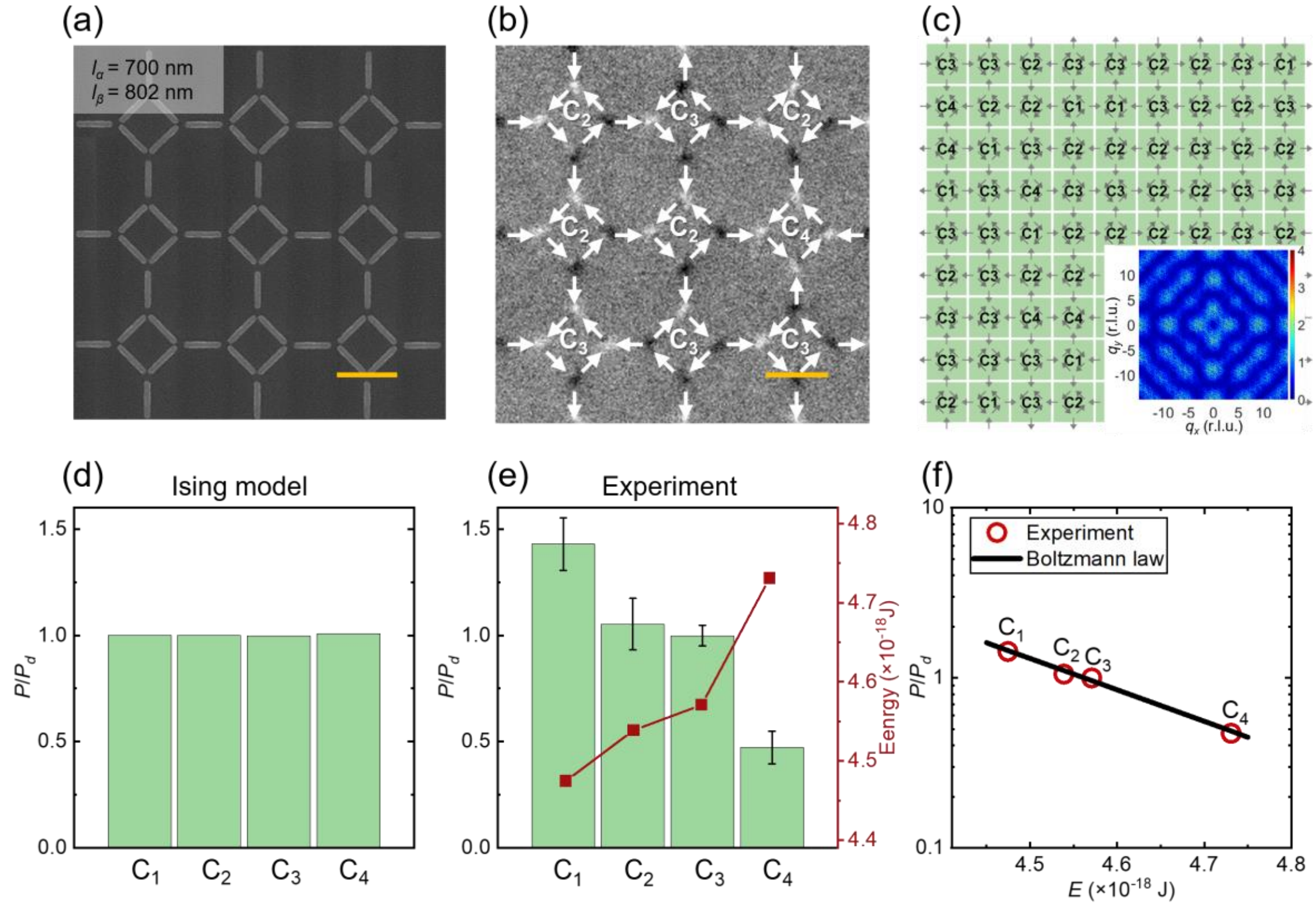

FIG. 2. Energetic relaxation and Boltzmann statistics in the spin-foam phase. (a) Scanning electron microscopy (SEM) image of a square-kagome ASI array, which corresponds to the spin-foam phase. The lattice constant is indicated in the upper left corner. Scale bars, 1 μm. (b) Magnetic force microscopy (MFM) image corresponding to (a), showing spin-foam phase of type C plaquettes. Arrows indicate the spin configurations. Scale bars, 1 μm. (c) Plaquette spin configurations from the boxed regions in Fig. S8h, showing a spin-foam state dominated by type-C plaquettes. Inset: Experimental magnetic structure factor map exhibiting diffuse scattering characteristic of the spin-foam phase. (d) and (e) Degeneracy-normalized plaquette populations, $P/P_d^i$, from the Ising-model simulation (d) and experiment (e), where the degeneracy populations are $P_d^{C1} = P_d^{C4}$ = 4/32=0.125, $P_d^{C2}$ = 8/32=0.25, and $P_d^{C3}$ = 16/32=0.5, consistent with an uncorrelated gas of $\beta$ spins. The red-square curve in (e), plotted on the right axis, indicates plaquette energies calculated from micromagnetic simulations including long-range dipolar interactions. Experimental error bars represent mean ± s.d. from two measurements. (f) Boltzmann-law fit to the degeneracy-normalized populations, $P/P_d^i$, of the four type-C subclasses, $C_1$-$C_4$, plotted as a function of their micromagnetic energies ($E_i$). The red circles show experimental values obtained from (e), and the black solid line shows the Boltzmann-law prediction, $P/P_d^i = e^{-E_i/(k_BT)}$. The vertical axis is plotted on a logarithmic scale.

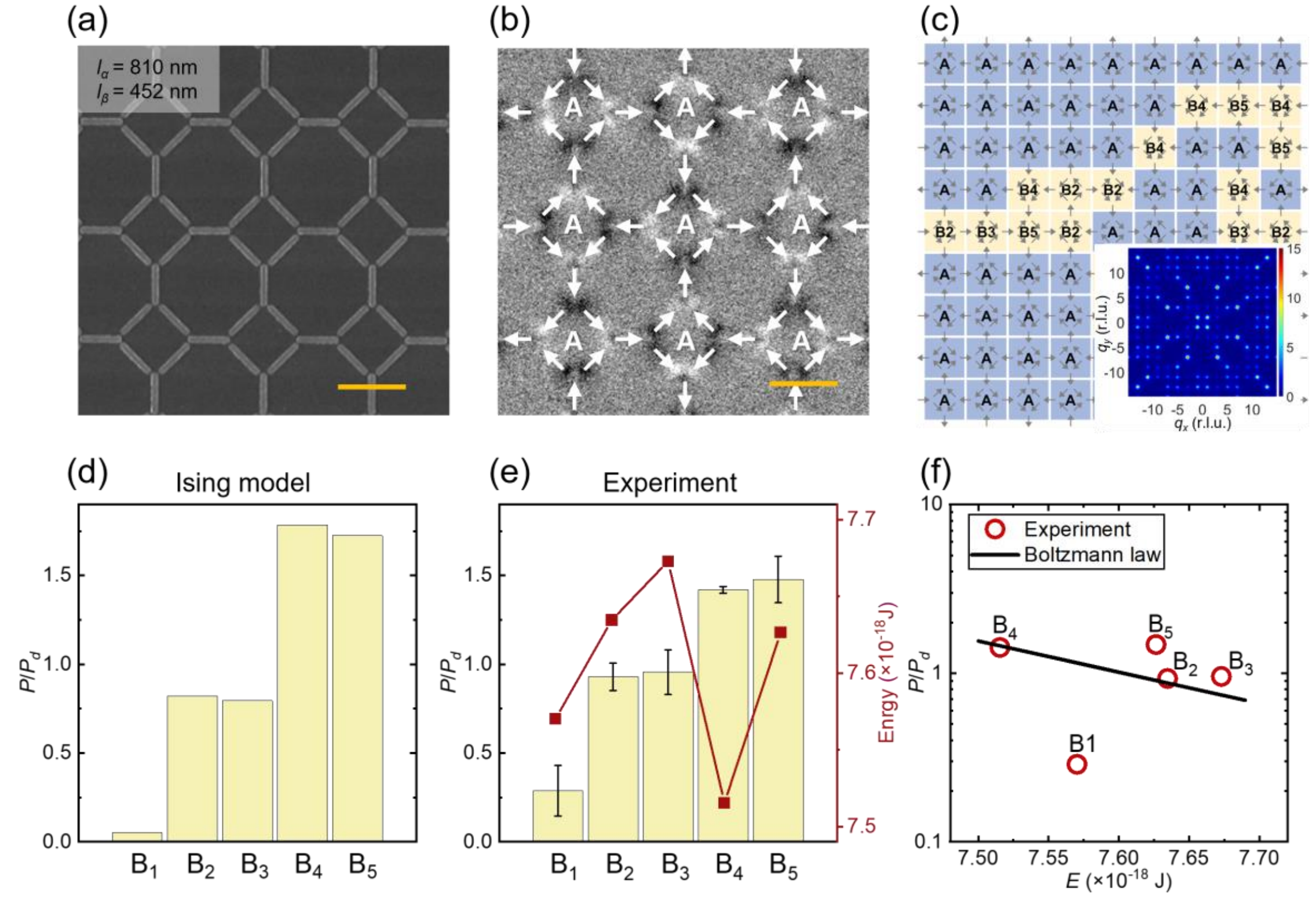

FIG. 3. Breakdown of Boltzmann statistics and kinetically constrained relaxation in the spin-crystal phase. (a) Scanning electron microscopy (SEM) image of a square-kagome ASI array, which corresponds to the spin-crystal phase. The lattice constant is indicated in the upper left corner. Scale bars, 1 μm. (b) Magnetic force microscopy (MFM) image corresponding to (a), showing spin-crystal phase of type A plaquettes. Arrows indicate the spin configurations. Scale bars, 1 μm. (c) Plaquette spin configurations from the boxed regions in Fig. S8a, showing a spin-crystal phase composed of two-fold-degenerate ordered domains of type-A plaquettes separated by domain walls made of type-B plaquette excitations, $B_1$-$B_5$. Inset: Experimental magnetic structure factor map showing sharp Bragg peaks characteristic of the spin-crystal phase. (d) and (e) Degeneracy-normalized plaquette populations, $P/P_d$ , from Ising-model Monte Carlo simulation (d) and from experiment (e), where the degeneracy populations are $P_d^{B1}$ =8/48, $P_d^{B2}$ =16/48, $P_d^{B3}$ =8/48, $P_d^{B4}$ =8/48, and $P_d^{B5}$ =8/48. In both cases, the normalized populations separate into three distinct groups: {$B_1$}, {$B_2$, $B_3$}, and {$B_4$, $B_5$}. Long-range dipolar interactions lift the degeneracy, with the corresponding excitation energies (obtained from micromagnetic simulation) shown by the red square curve in (e) (right axis). Experimental error bars represent mean ± s.d. from two measurements. (f) Comparison between the degeneracy-normalized populations, $P/P_d^i$, of types $B_1$-$B_5$, and the Boltzmann-law prediction. The red circles show experimental values obtained from (e), and the black solid line shows the best-fit Boltzmann-law prediction, $P/P_d^i = e^{-E_i/(k_BT)}$. The vertical axis is plotted on a logarithmic scale. The clear discrepancy between the experimental data and the Boltzmann-law prediction indicates a breakdown of Boltzmann statistics.

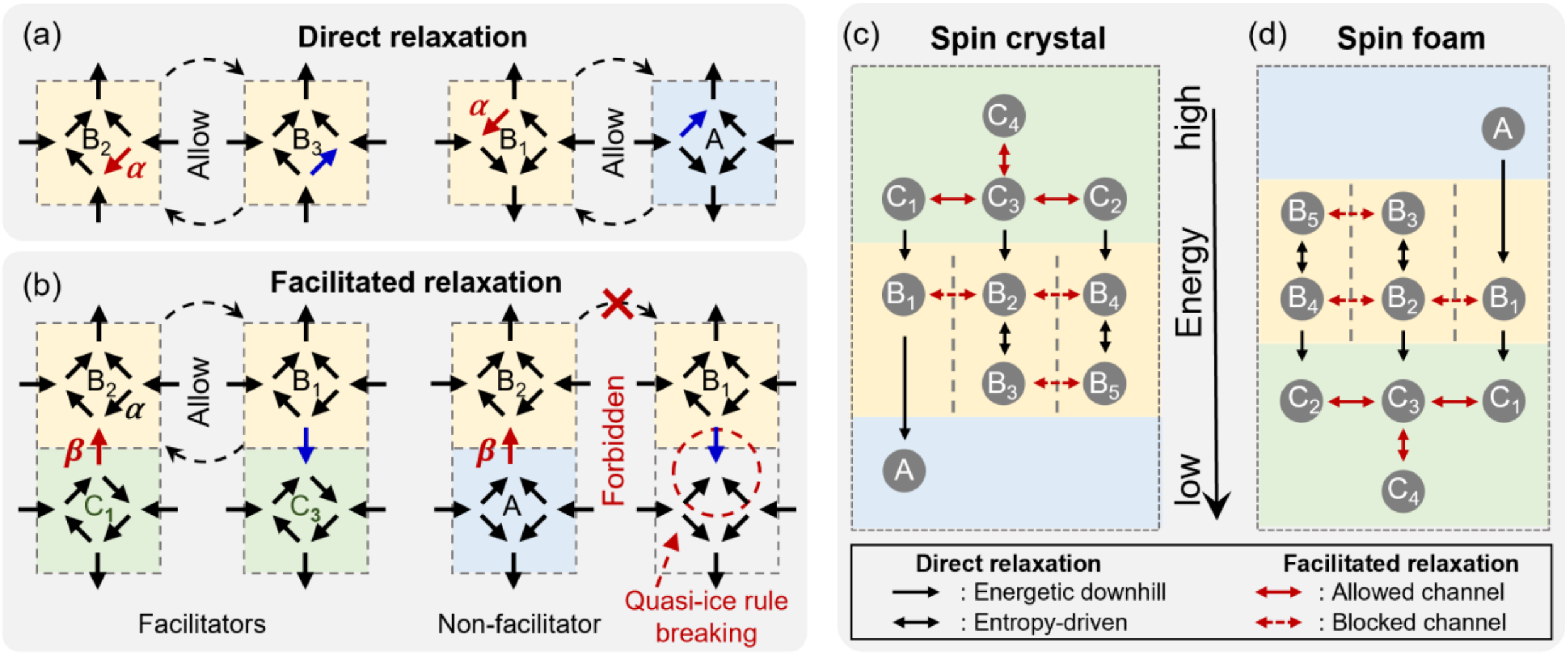

FIG. 4. Phase-dependent accessibility of facilitated relaxation pathways. (a) Examples of direct relaxations induced by flipping an $\alpha$ spin within a plaquette. Such transitions (e.g., $B_2$ to $B_3$, left; $B_1$ to A, right) modify only the local plaquette configuration and do not affect the energetics or configurations of neighboring plaquettes. (b) Examples of facilitated relaxations induced by flipping a $\beta$ spin that connects two neighboring plaquettes. For instance, the transition from $B_2$ to $B_1$ proceeds through a facilitated pathway that is allowed when the neighboring plaquette acts as a specific *facilitator*, such as a type-C plaquette (left panels), but is blocked when the neighboring plaquette is a *non-facilitator*, such as a type-A plaquette (right panels), as the latter transition would violate the quasi-ice rule at the shared vertex. (c) and (d) Relaxation networks connecting the ten topologically distinct plaquette groups for the spin-crystal phase (c) and the spin-foam phase (d). Black arrows denote direct relaxations, one-way arrows indicate energetically downhill relaxations, and two-way arrows indicate entropy-driven relaxations between degenerate configurations in the Ising model. Red arrows denote facilitated relaxation channels. Dashed red arrows in (c) denote facilitated channels that are blocked by the type-A (*non-facilitator*) ground-state background of the spin-crystal phase, thereby separating the type-B domain-wall excitations into three dynamically disconnected groups, $\{B_1\}$, $\{B_2, B_3\}$, and $\{B_4, B_5\}$, within which only direct and entropy-driven relaxations are accessible. Solid red arrows in (d) indicate facilitated channels that are accessible in the type-C (*facilitator*) ground-state background of the spin-foam phase. Detailed transitions between the plaquette configurations are shown in Fig. S11.

**Acknowledgements**

This work was supported by the Quantum Science and Technology-National Science and Technology Major Project (Grant No. 2024ZD0301300), the National Key R&D (Grant Nos. 2021YFA0718802, 2024YFA1408900), the National Natural Science Foundation of China (Grant Nos. [62274086, 62288101, 62101243, 92565201, 12504137]), the Fundamental Research Funds for the Central Universities (Grant No. 0210-14380277), the China Postdoctoral Science Foundation (Grant No. 2024M751370, 2025T180930), the Jiangsu Funding Program for Excellent Postdoctoral Talent (Grant No. 2023ZB534), Natural Science Foundation of Jiangsu Province (Grant No. BK20241223). The work of C.N. was carried out under the auspices of the U.S. DoE through the Los Alamos National Laboratory, operated by Triad National Security, LLC (Contract No. 229892333218NCA000001) and funded by a grant from the DOE-LDRD office.

# Supporting Information:

**Facilitated Relaxation Overrides Energetic Pathways in a Multiphase Square-Kagome Artificial Spin Ice**

Wen-Cheng Yue[1,2,#], Zixiong Yuan[1,2,†,#], Peiyuan Huang[1,2], Ziao Wang[1,2], Hangyu Liu[3], Tianyu Li[1,2], Nuo-Zhou Yang[1,2], Jingyue Gao[1,2], Chen Ku[1,2], Shuo Song[1,2], Jiyao Shang[1,2], Yang-Yang Lyu[1,2], Yihuang Xiong[1], Xuecou Tu[1,2], Liang He[1], Xun Cao[1], Ying Dong[3,*], Huabing Wang[1,2,*], Peiheng Wu[1,2], Cristiano Nisoli[4,*], and Yong-Lei Wang[1,2,*]

[1] *Research Institute of Superconductor Electronics, State Key Laboratory of Spintronics Devices and Technologies, Nanjing University, Nanjing, China*

[2] *Purple Mountain Laboratories, Nanjing, China*

[3] *College of Metrology Measurement and Instrument, China Jiliang University, Hangzhou, China*

[4] *Theoretical Division and Center for Nonlinear Studies, Los Alamos National Laboratory, Los Alamos, NM, USA*

† Present address: Sichuan Province Engineering Research Center for Broadband Microwave Circuit High Density Integration, Chengdu, China

# Authors contributed equally

* Correspondence to: yingdong@cjlu.edu.cn; hbwang@nju.edu.cn; cristiano@lanl.gov; yongleiwang@nju.edu.cn

## Geometric structure of the square-kagome ASI and plaquette classification

The square-kagome artificial spin ice (ASI) consists of nanomagnets placed at the vertices of a square-kagome lattice, as shown in Fig. S1.

As illustrated in Fig. 1a, each unit square plaquette contains eight nanomagnets, which give rise to $2^8 = 256$ possible spin configurations. Among these, 82 configurations satisfy the quasi-ice rule at all four vertices and therefore correspond to low energy states, as shown in Fig. S2. In Fig. S2, the plaquette spin configurations are classified according to the arrangements of the α and β spins, each of which admits four symmetrical configurations, leading to $4^2 = 16$ topologically distinct configuration groups for each plaquette. Of these, only ten configuration groups satisfy the quasi-ice rule; these are highlighted in Fig. S2 and Fig. 1c.

Based on the nearest-neighbor interaction energies of frustrated spin pairs, the 82 low-energy plaquette configurations can be further grouped into three classes—A, B (I-B and II-B), and C —highlighted in blue, orange, and green, respectively (Fig. S2). These classes are determined solely by the configurations of the α spins, as listed in Fig. S2. As mentioned in the main text, the interaction energies of the three classes are $4J_{\alpha\alpha}$, $2J_{\alpha\alpha}+2J_{\alpha\beta}$, and $4J_{\alpha\beta}$ (the pairing energy of the head-to-tail spins is defined as zero), with degeneracies of 2, 48 and 32, respectively. A unique property of the square-kagome ASI is that each degenerate plaquette type contains multiple topologically distinct (symmetrically different) configurations. For example, the 48-fold degenerate type-B plaquettes consist of five topologically distinct configurations ($B_1$-$B_5$), while the 32-fold degenerated type-C plaquettes consist of four topologically distinct configurations ($C_1$-$C_4$). The degeneracy *d* of each group corresponds to the number of topologically equivalent configurations related by rotation, mirror, and/or all-spins inversion operations (see Fig. S3).

## Sample fabrication

The details of sample fabrication have been described previously [27]. All nine square–kagome ASI arrays were fabricated on the same 1 cm × 1 cm silicon substrate with a 200 nm silicon nitride layer. A bilayer resist consisting of PMMA 495 and PMMA 950 was spin-coated for electron-beam lithography. After patterning, a 15-nm-thick permalloy ($Ni_{0.8}Fe_{0.2}$) film was deposited by electron-beam evaporation at a rate of 0.3 Å/s. To prevent oxidation, a 3-nm-thick aluminum capping layer was deposited on top. Each fabricated array contains approximately $2 \times 10^4$ individual nanomagnets.

Representative scanning electron microscopy (SEM) images of the square-kagome ASI are shown in Fig. S5.

**Demagnetization protocol**

The sample with nine square–kagome ASI arrays was mounted on a motor rotating at 2000 rpm. An oscillating in-plane magnetic field with a sinusoidal waveform and a period of 40 s was applied. The amplitude of the magnetic field was gradually reduced from 1000 G (well above the saturation field of the nanomagnets) to zero over a total duration of 72 hours.

**Micromagnetic simulations**

We define the interaction energy between nearest-neighbor two α magnets as $J_{\alpha\alpha}$ ($E_{\alpha\alpha}^{\text{high}} - E_{\alpha\alpha}^{\text{low}}$), and that between nearest-neighbor α and β magnets as $J_{\alpha\beta}$ ($E_{\alpha\beta}^{\text{high}} - E_{\alpha\beta}^{\text{low}}$). The interaction energy is defined as the difference in magnetostatic energy between the high-energy and low-energy configurations of a given magnet pair. Accordingly, the interaction energy of a nanomagnet pair with a head-to-tail configuration is set to zero.

Micromagnetic simulations were performed using the MuMax3 package [49]. The material parameters for permalloy ($Ni_{0.8}Fe_{0.2}$) were set as follows: exchange constant A = 1.3 × $10^{-11}$ J/m, saturation magnetization $M_s$ = 8.6 × $10^5$ A/m, and Gilbert damping constant α = 0.01. The simulation cell size was 2 nm × 2 nm × 2 nm.

The square-kagome ASI admits ten topologically distinct plaquette configuration classes. Based on their nearest-neighbor interaction energetics, $E_A = 4J_{\alpha\alpha}$, $E_B = 2J_{\alpha\alpha} + 2J_{\alpha\beta}$, $E_C = 4J_{\alpha\beta}$, these classes are categorized into three plaquette types: A, B, and C (Fig. 1c). We calculated the coupling strengths $J_{\alpha\alpha}$ and $J_{\alpha\beta}$ for all investigated lattice constants using the same framework. When second-nearest-neighbor interactions are included, the energies of the ten classes are modified accordingly, as summarized in Table S1.

**Monte Carlo simulations**

Monte Carlo simulations were performed using a single-spin-flip algorithm on a square-kagome lattice comprising 30×30 sites with free boundary conditions. The system is described by the Hamiltonian $\mathrm{H} = -\sum_{\langle i,j \rangle} J_{ij} S_i S_j$, where $S_i$ and $S_j$ are Ising variables on sites $i$ and $j$, respectively, and $J_{ij}$ denotes the coupling strength between nearest-neighbor sites. Two coupling values, $J_{\alpha\alpha}$ and $J_{\alpha\beta}$, are considered.

To reach the ground state, a thermal annealing simulation protocol was employed in which the temperature was linearly decreased from $T = 150$ to $T = 0$ over $5\times10^4$ Monte Carlo steps for each simulated sample. This procedure mimics a finite-duration thermal annealing process and was applied consistently across all simulations.

**Magnetic spin structure factor.**

The magnetic spin structure factor was calculated by following established methods. We define the spin component $\mathbf{S}^{\perp}$ as:

$$\vec{\mathrm{S}}^{\perp} = \vec{\mathrm{S}} - (\hat{\vec{\mathrm{q}}} \cdot \vec{\mathrm{S}})\, \hat{\vec{\mathrm{q}}}$$

where $\hat{\vec{\mathrm{q}}}$ is the unit scattering vector: $\hat{\vec{\mathrm{q}}} = \dfrac{\vec{\mathrm{q}}}{\|\vec{\mathrm{q}}\|}$

For each vector $\boldsymbol{q} = (q_x, q_y)$, the magnetic spin structure factor intensity $I(\boldsymbol{q})$ is given by:

$$I(\vec{q}) = \frac{1}{N} \sum_{(i,j=1)}^{N} \vec{S}_i^{\perp} \cdot \vec{S}_j^{\perp} \exp(i\vec{q} \cdot (\vec{r}_i - \vec{r}_j))$$

The magnetic spin structure factor was calculated over the interval $(q_x, q_y) = [-14.9\pi, -14.9\pi] - [14.9\pi, 14.9\pi]$ in 513×513 steps.

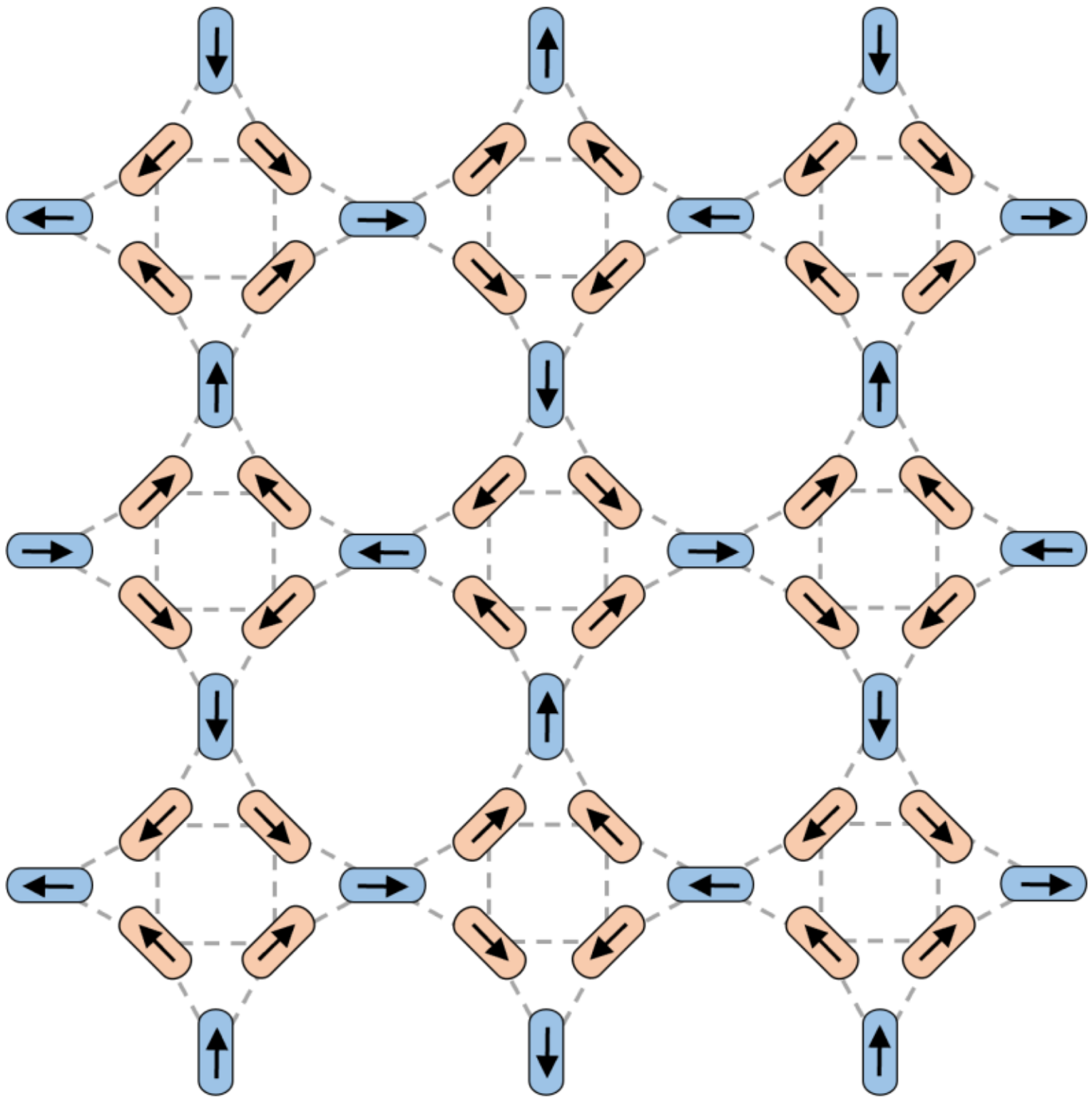

FIG. S1. Schematic of a square-kagome artificial spin ice. Nanomagnets are positioned at the vertices of a square-kagome lattice (gray-dash lines).

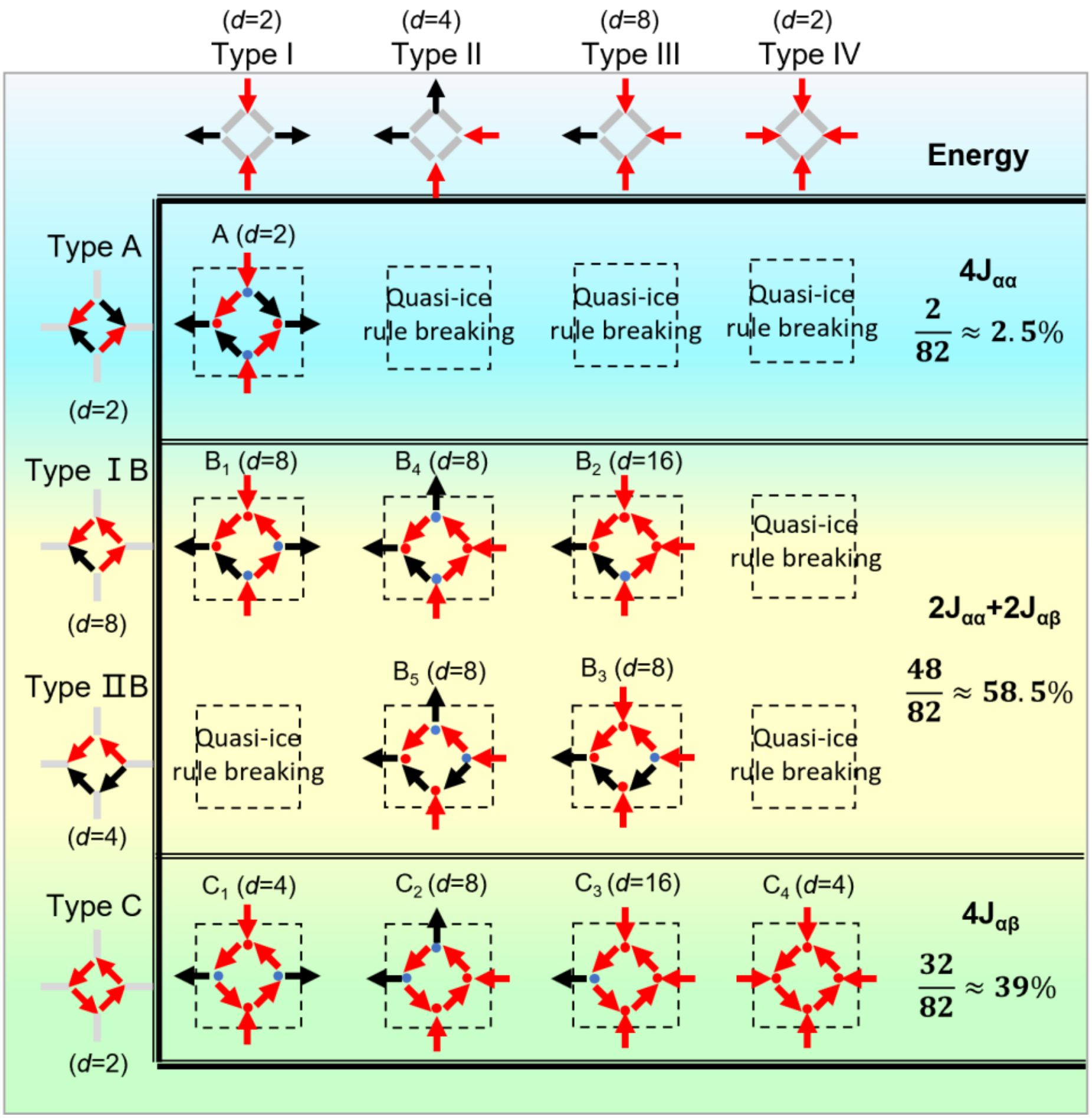

FIG. S2. Classification of the ten topologically distinct low-energy plaquette configurations. These configurations satisfy the quasi-ice rule at all four vertices and are classified based on the arrangements of α (rows) and β (columns) spins. The degeneracy *d* represents the number of topologically equivalent configurations within a group related by rotation, mirror, and/or global spin inversion operations (see Fig. S3).

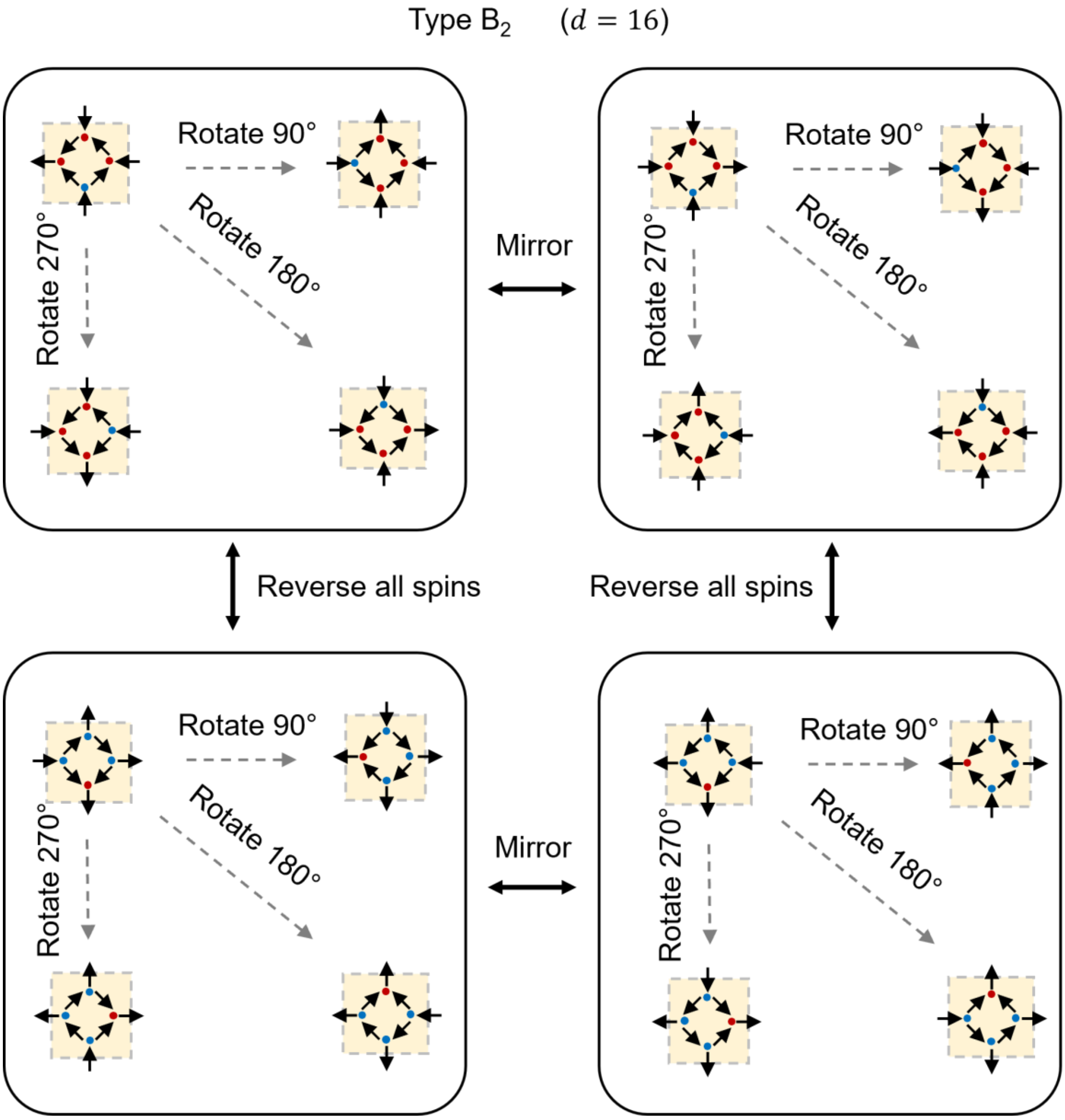

FIG. S3. Example of topologically equivalent spin configurations. Illustration of the 16 topologically equivalent configurations of type-$B_2$ that can be transformed into one another via symmetry operations (rotation, mirror, or global spin inversion). Note that configurations from different topological groups cannot be transformed into one another through these operations.

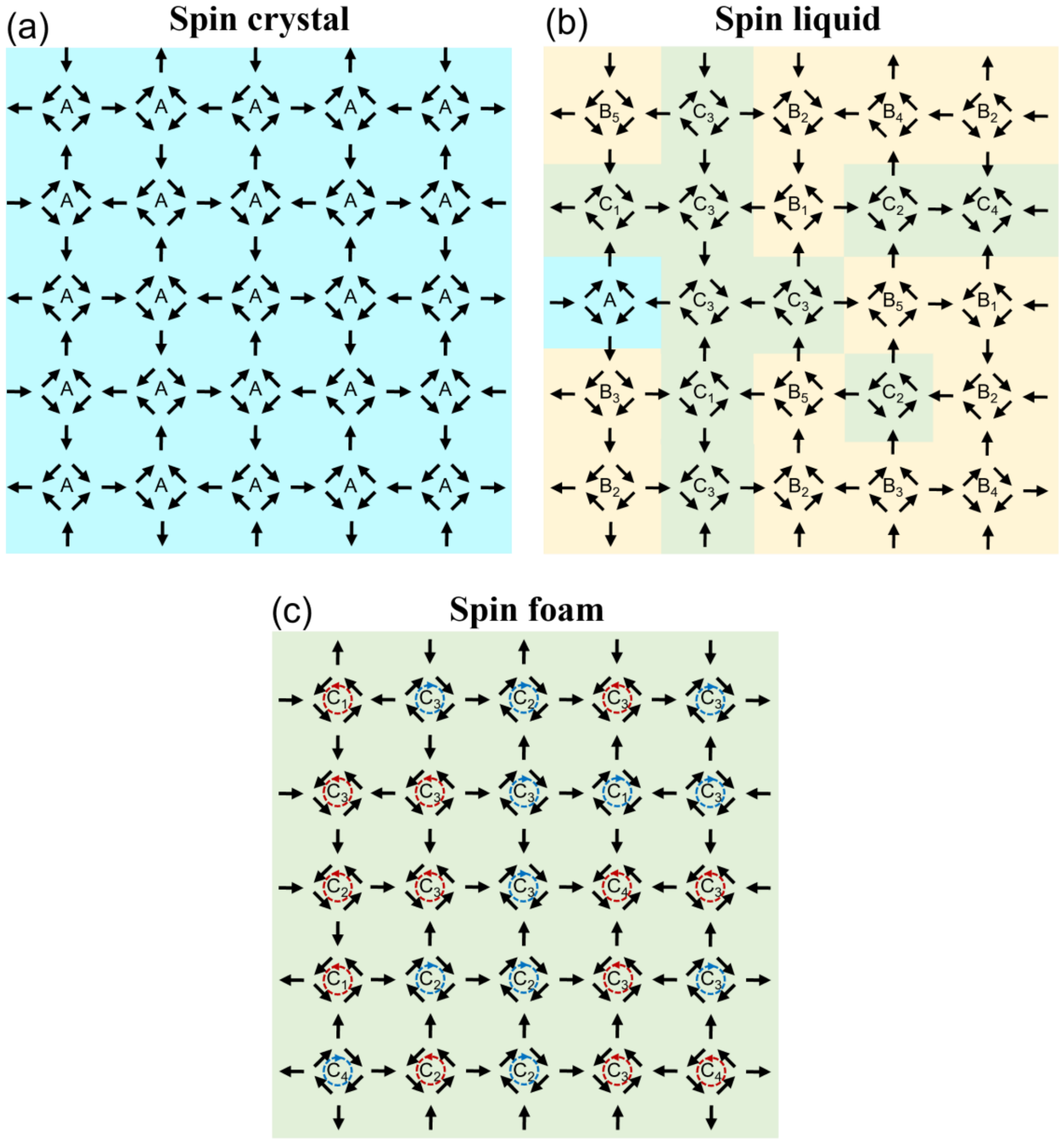

FIG. S4. Spin configurations of the theoretically predicted low-temperature phases. (a) A two-fold degenerate spin crystal composed entirely of type A plaquettes. (b) A spin-liquid state where all 82 quasi-ice-rule-satisfying configurations (types A, $B_1$-$B_5$, and $C_1$-$C_4$) are degenerate. (c) A spin-foam phase dominated by types $C_1$-$C_4$ configurations, where $\alpha$ magnets form toroidal moments while $\beta$ magnets remain frustrated and highly dynamic (Video 2).

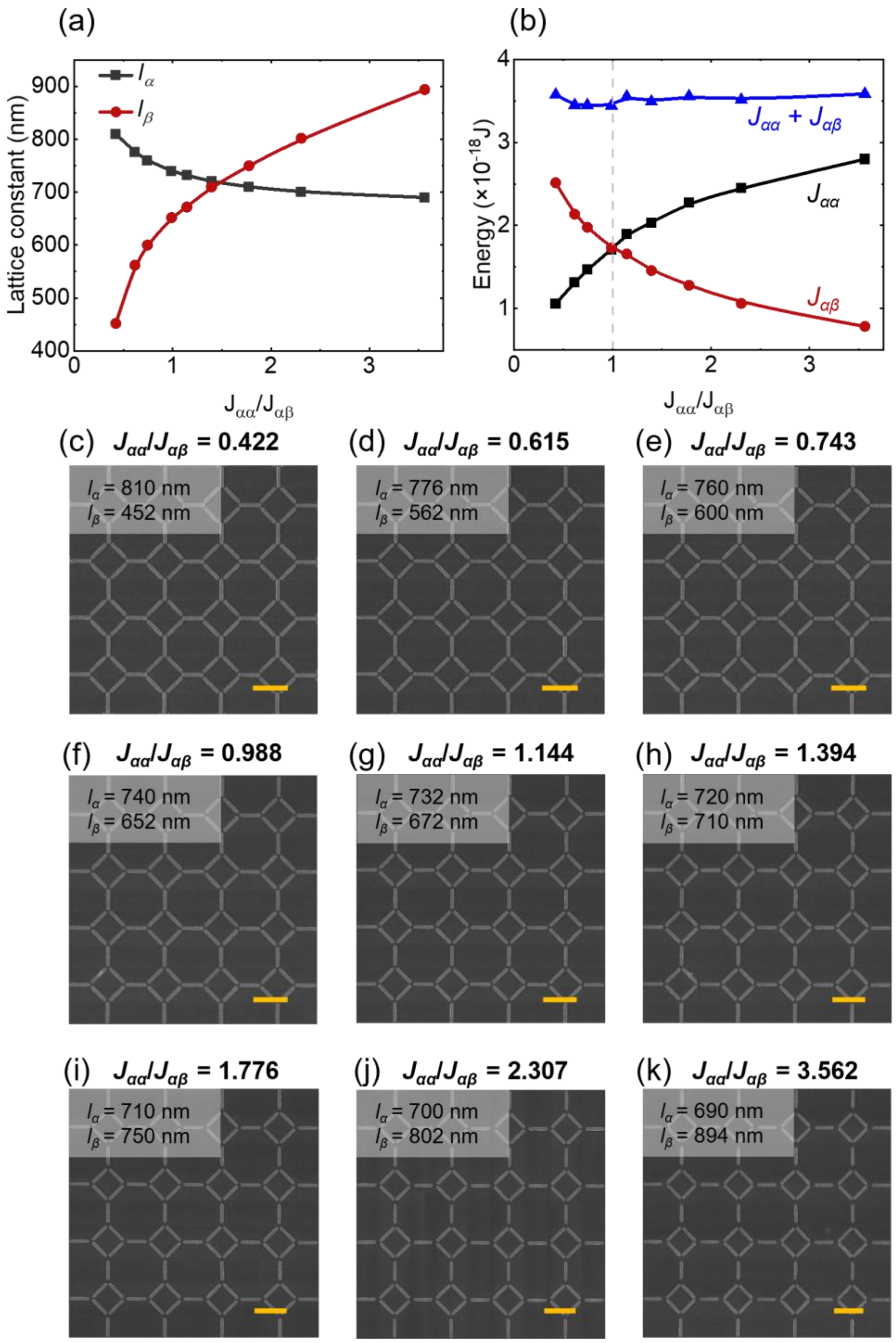

FIG. S5. Sample parameters and SEM images. (a) Lattice constants for the nine fabricated samples plotted against the coupling ratios $J_{\alpha\alpha}/J_{\alpha\beta}$ derived from the energies in (b). (b), Interaction energies calculated via micromagnetic simulations. The sum of the nearest-neighbor coupling strengths, $J_{\alpha\alpha}+J_{\alpha\beta}$, is designed to remain approximately constant. Representative micromagnetic configurations used to determine these interaction strengths are provided in Fig. S13. (c)-(k), Representative SEM images of the nine samples with various coupling ratios. Scale bars, 1 μm.

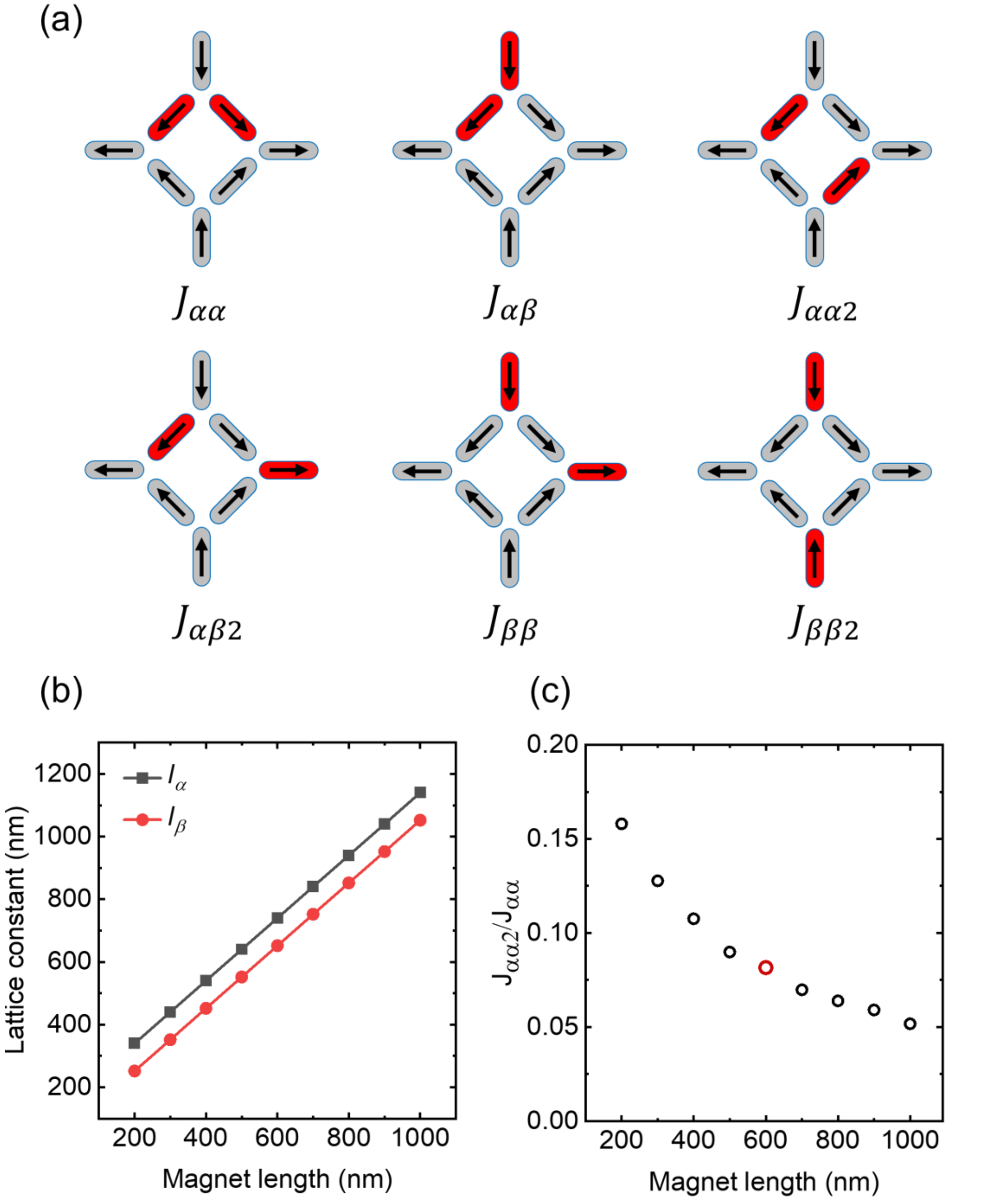

FIG. S6. Coupling ratios between nearest-neighbor and next-nearest-neighbor interactions as a function of nanomagnet length. (a) Illustration of nearest-neighbor ($J_{\alpha\alpha}$ and $J_{\alpha\beta}$) and next-nearest-neighbor ($J_{\alpha\alpha2}$, $J_{\alpha\beta2}$, $J_{\beta\beta}$, and $J_{\beta\beta2}$) coupled nanomagnets/spins (highlighted in red). (b) The two lattice constants vary with nanomagnet length, maintaining a constant gap at the vertices. (c) Energy ratio of next-nearest-neighbor $J_{\alpha\alpha2}$ to nearest-neighbor $J_{\alpha\alpha}$ as a function of nanomagnet length. A ratio of 8% with a nanomagnet length of 600 nm used in this study is highlighted in red.

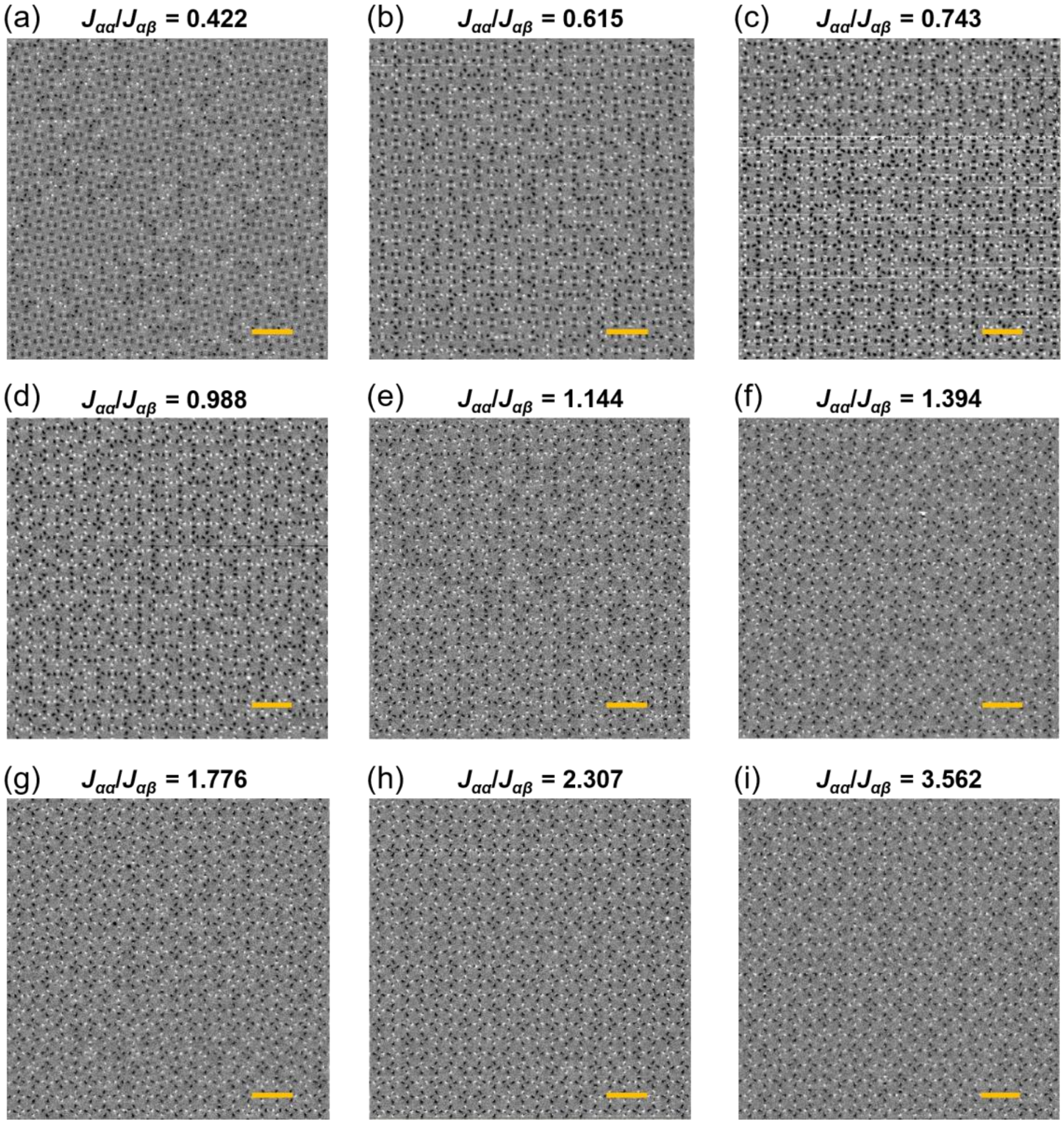

FIG. S7. Magnetic Force Microscopy (MFM) images. (a)-(i), MFM images of the nine samples across various $J_{\alpha\alpha}/J_{\alpha\beta}$ ratios, captured after demagnetization. Scale bars, 5 μm.

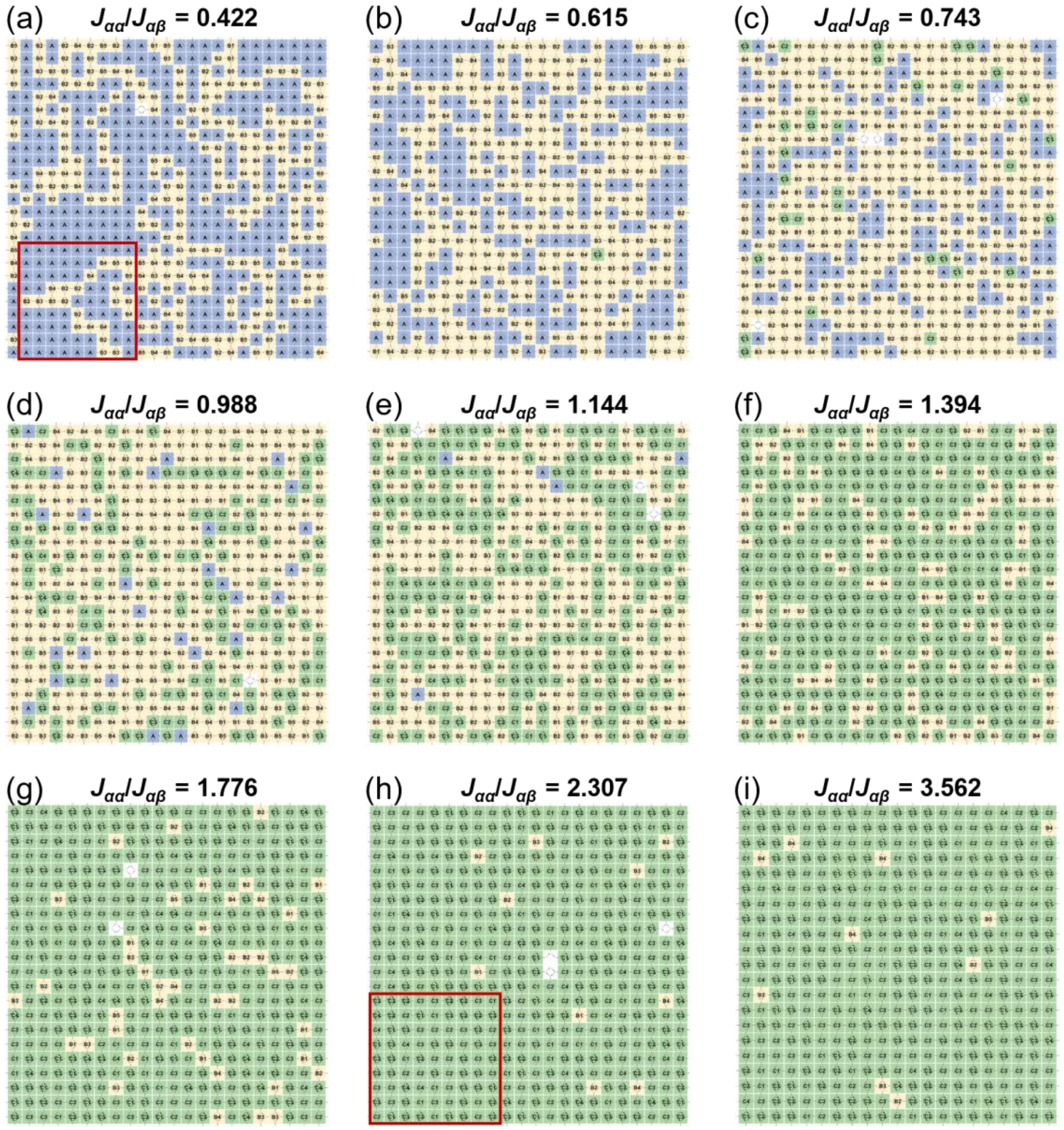

FIG. S8. Extracted plaquette configurations. Spin configurations derived from the MFM images in Fig. S7. The boxed regions in (a) and (h) mark the areas shown in Figs. 3c and 2c, respectively.

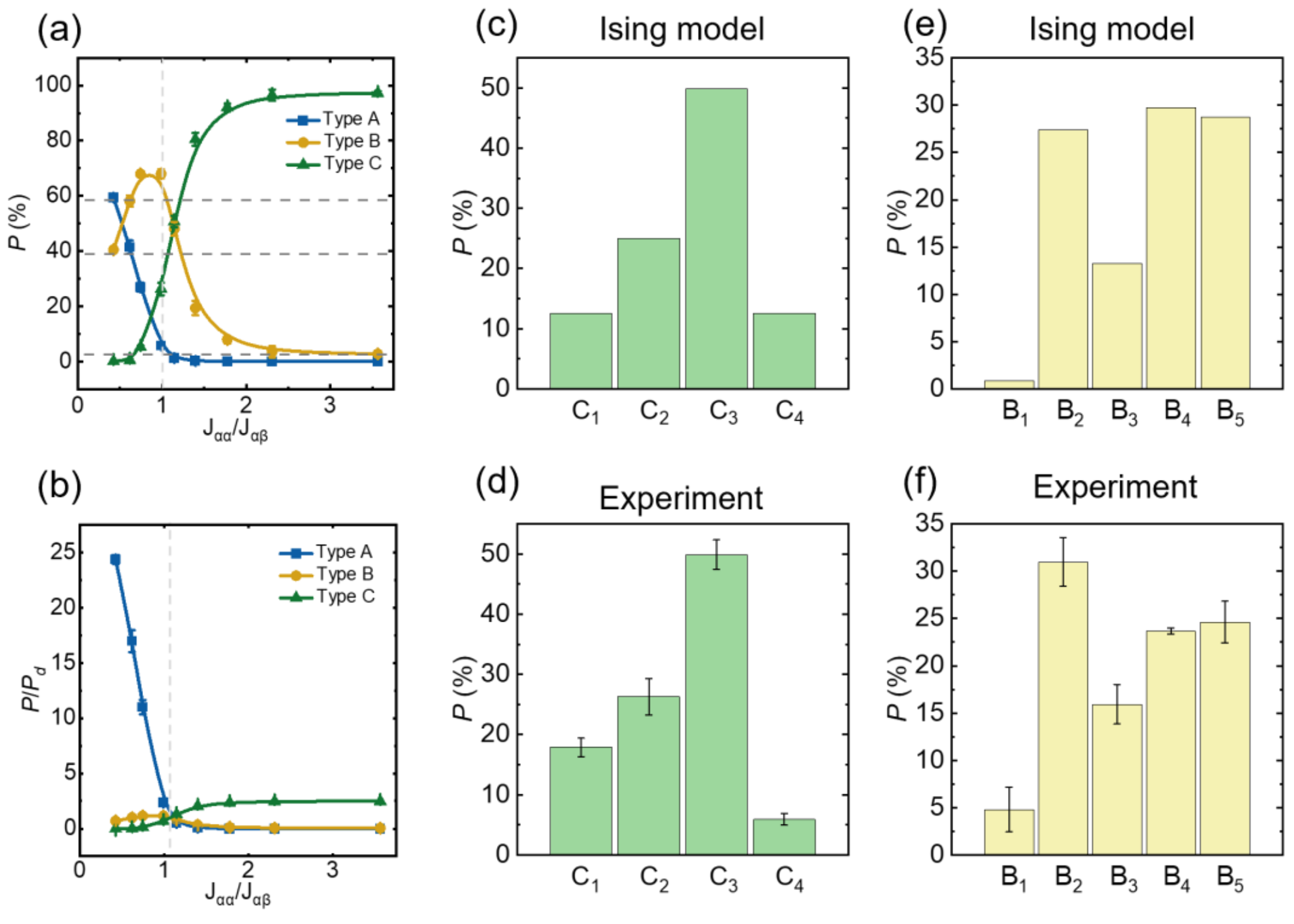

FIG. S9. Experimental statistics of the plaquette populations. (a) Raw plaquette populations $P$ for types A, B, and C as a function of the coupling ratio $J_{\alpha\alpha}/J_{\alpha\beta}$, extracted from MFM images after demagnetization (Fig. S7). Horizontal dashed lines indicate the corresponding degeneracy populations $P_d^A = \frac{2}{82}$, $P_d^B = \frac{48}{82}$, and $P_d^C = \frac{32}{82}$. (b) Experimental statistics of the degeneracy-normalized plaquette populations, $P/P_d^i$, as a function of the coupling ratio $J_{\alpha\alpha}/J_{\alpha\beta}$. (c) and (d) Histograms of the populations of the four plaquette types ($C_1$–$C_4$) in the spin-foam phase, obtained from (c) Monte Carlo simulations of the Ising model and (d) magnetic force microscopy (MFM) measurements at a coupling ratio of $J_{\alpha\alpha}/J_{\alpha\beta} = 2.307$. (e) and (f) Histograms of the populations of the five subclasses ($B_1$–$B_5$) of type-B domain-wall excitations, obtained from (e) Monte Carlo simulations of the Ising model and (f) experiments at a coupling ratio of $J_{\alpha\alpha}/J_{\alpha\beta} = 0.422$.

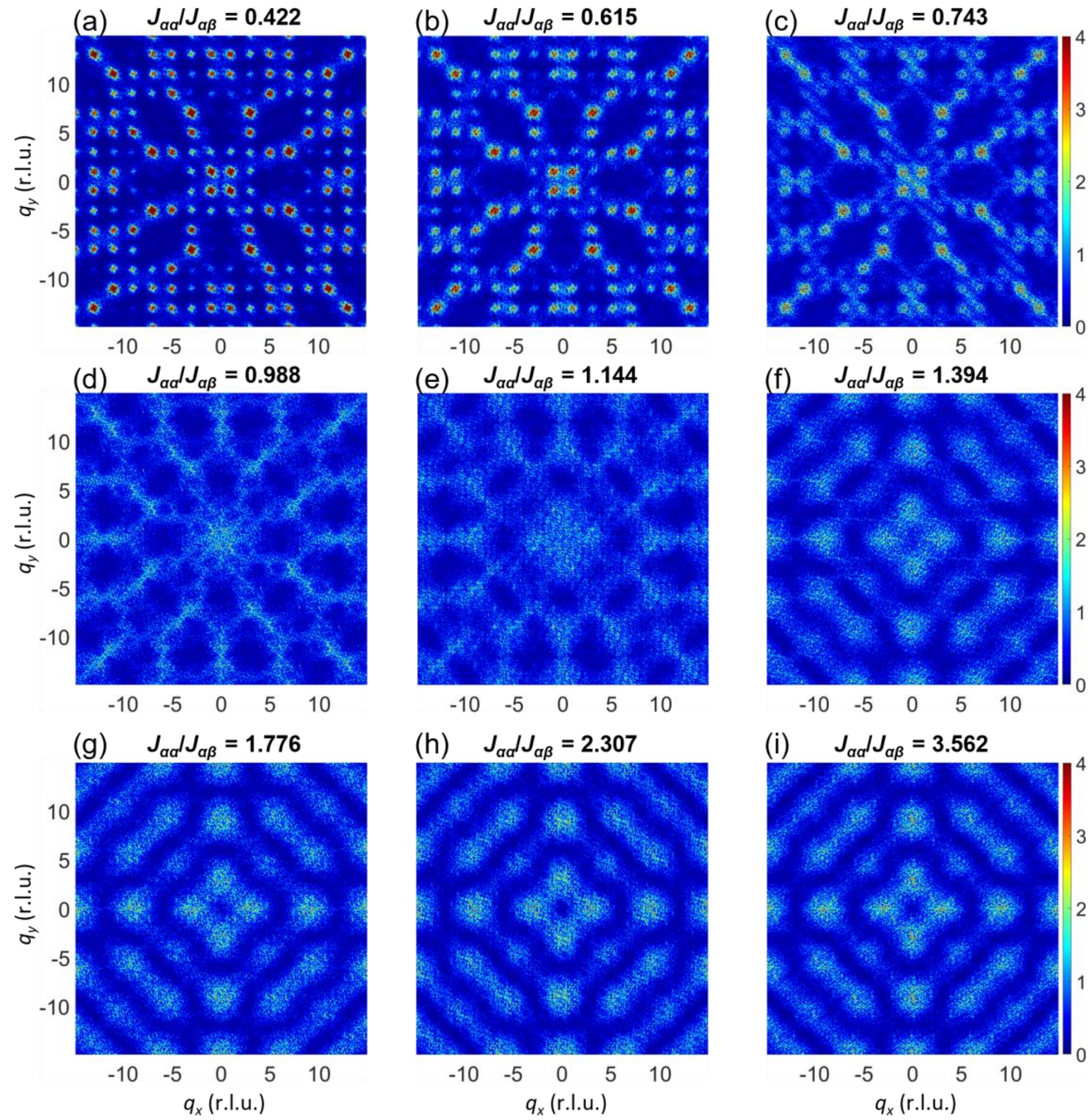

FIG. S10. Experimental Magnetic Structure Factor (MSF) maps. (a)-(i) Calculated MSF maps for all experimental samples, showing the evolution from sharp Bragg peaks (spin-crystal phase) to diffuse and structured features (spin-liquid and spin-foam phases).

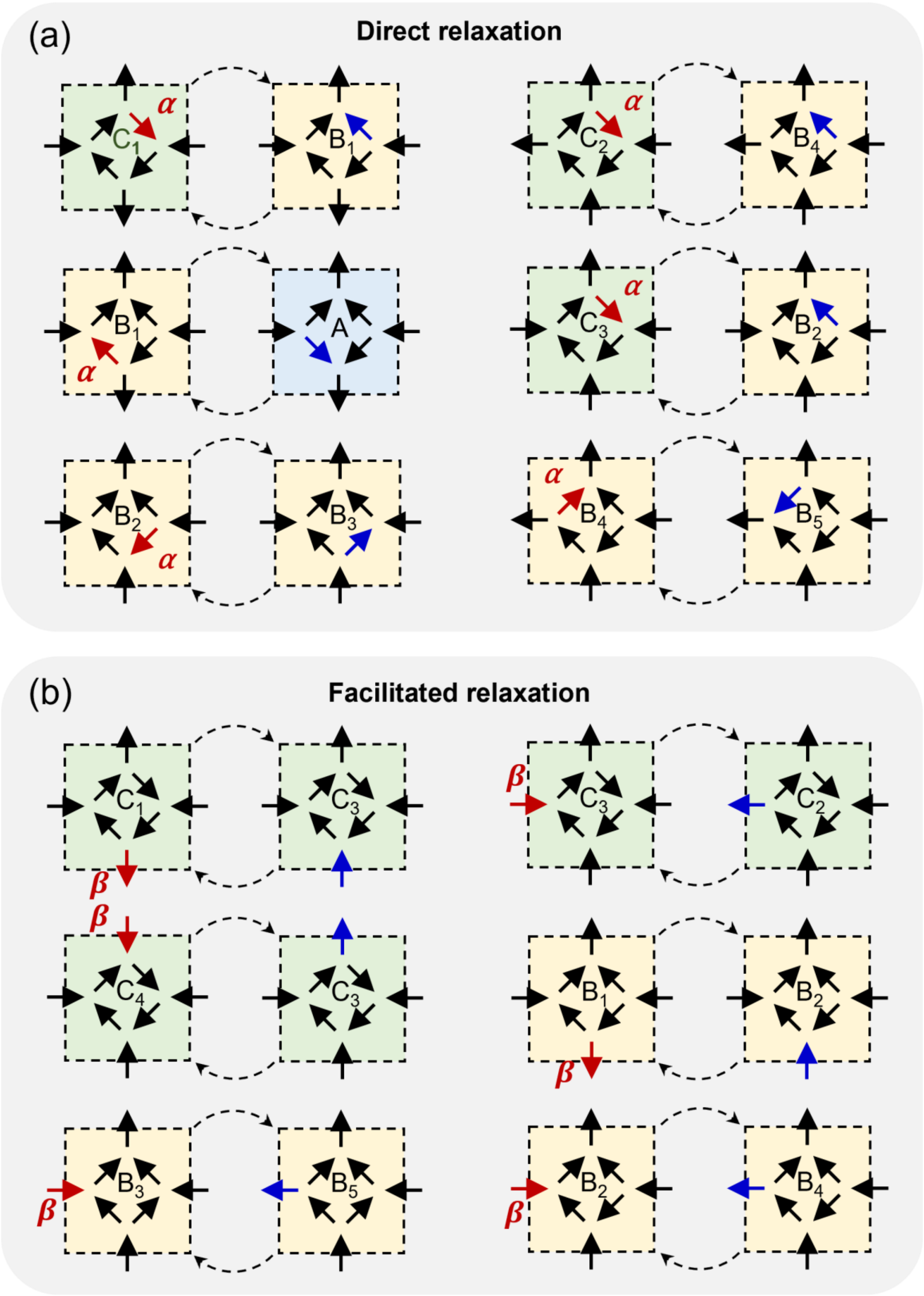

FIG. S11. Transitions between the ten topologically distinct plaquette configurations. (a) Spin flips associated with direct relaxation pathways. (b) Spin flips associated with facilitated relaxation pathways.

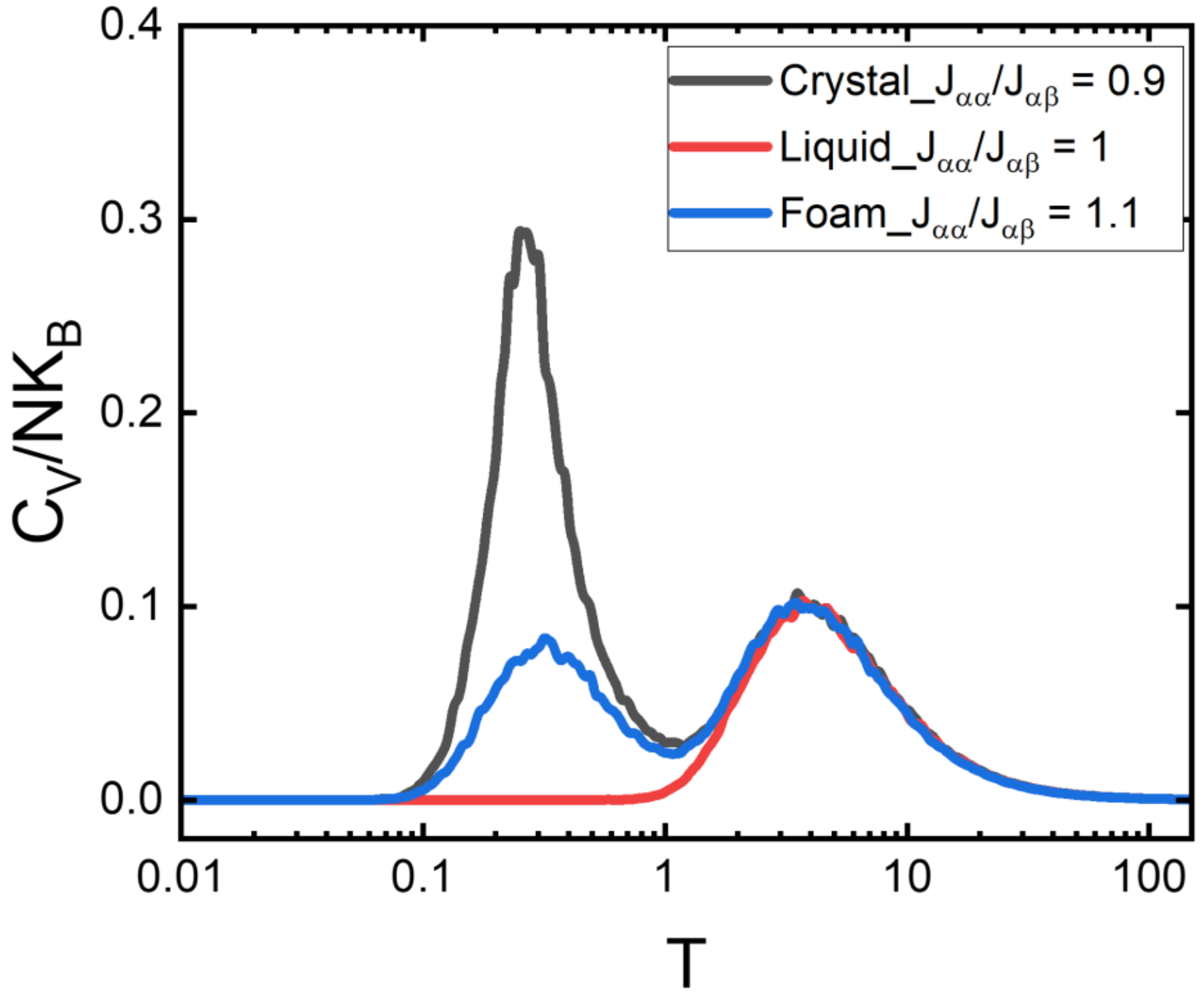

FIG. S12. Temperature-dependent heat capacity from Monte Carlo simulations. Three representative heat capacity curves calculated for $J_{\alpha\alpha}/J_{\alpha\beta}$=0.9, 1, and 1.1.

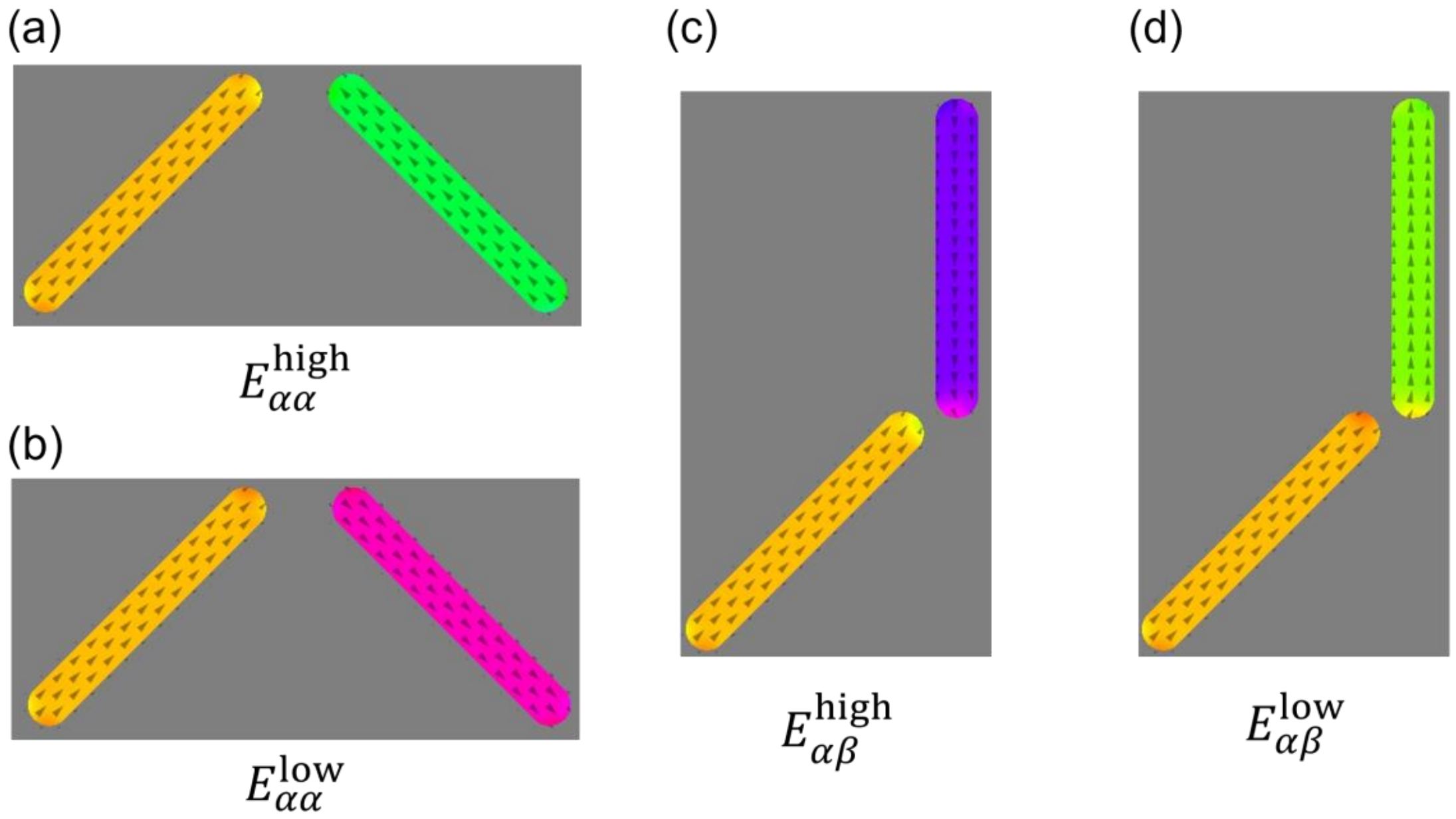

FIG. S13. Representative micromagnetic configurations used to calculate the nearest-neighbor interaction energies. (a,b) Magnetization distributions of an α–α nanomagnet pair in the high-energy (a) and low-energy (b) configurations. (c,d) Corresponding magnetization distributions of an α–β nanomagnet pair in the high-energy (c) and low-energy (d) configurations. The configurations were obtained from MuMax3 simulations using the nanomagnet geometries considered in this study. The interaction strengths, $J_{\alpha\alpha} = E_{\alpha\alpha}^{\rm high} - E_{\alpha\alpha}^{\rm low}$ and $J_{\alpha\beta} = E_{\alpha\beta}^{\rm high} - E_{\alpha\beta}^{\rm low}$, are defined as the differences in magnetostatic energy between the corresponding high- and low-energy configurations, with the low-energy configurations taken as the reference. These interaction strengths determine the nearest-neighbor energy hierarchy of the plaquette configurations discussed in the main text.

| | | $J_{\alpha\alpha}/J_{\alpha\beta}$ = **0.422** | $J_{\alpha\alpha}/J_{\alpha\beta}$ = **0.988** | $J_{\alpha\alpha}/J_{\alpha\beta}$ = **2.307** |
|---|---|---|---|---|
| $E_A$ | $4J_{\alpha\alpha} + 8J_{\alpha\beta2} + 2J_{\beta\beta2}$ | 4.623 × $10^{-18}$J | 7.237 × $10^{-18}$J | 10.188 × $10^{-18}$J |
| $E_{B1}$ | $2J_{\alpha\alpha} + 2J_{\alpha\beta} + J_{\alpha\alpha2} + 6J_{\alpha\beta2} + 2J_{\beta\beta2}$ | 7.570 × $10^{-18}$J | 7.337 × $10^{-18}$J | 7.391 × $10^{-18}$J |
| $E_{B2}$ | $2J_{\alpha\alpha} + 2J_{\alpha\beta} + J_{\alpha\alpha2} + 4J_{\alpha\beta2} + 2J_{\beta\beta} + J_{\beta\beta2}$ | 7.635 × $10^{-18}$J | 7.371 × $10^{-18}$J | 7.505 × $10^{-18}$J |
| $E_{B3}$ | $2J_{\alpha\alpha} + 2J_{\alpha\beta} + 2J_{\alpha\alpha2} + 2J_{\alpha\beta2} + 2J_{\beta\beta} + J_{\beta\beta2}$ | 7.673 × $10^{-18}$J | 7.429 × $10^{-18}$J | 7.582 × $10^{-18}$J |
| $E_{B4}$ | $2J_{\alpha\alpha} + 2J_{\alpha\beta} + J_{\alpha\alpha2} + 2J_{\alpha\beta2} + 2J_{\beta\beta}$ | 7.515 × $10^{-18}$J | 7.253 × $10^{-18}$J | 7.392 × $10^{-18}$J |
| $E_{B5}$ | $2J_{\alpha\alpha} + 2J_{\alpha\beta} + 2J_{\alpha\alpha2} + 2J_{\alpha\beta2} + 2J_{\beta\beta}$ | 7.626 × $10^{-18}$J | 7.391 × $10^{-18}$J | 7.550 × $10^{-18}$J |
| $E_{C1}$ | $4J_{\alpha\beta} + 4J_{\alpha\beta2} + 2J_{\beta\beta2}$ | 10.295 × $10^{-18}$J | 7.160 × $10^{-18}$J | 4.475 × $10^{-18}$J |
| $E_{C2}$ | $4J_{\alpha\beta} + 4J_{\alpha\beta2} + 2J_{\beta\beta}$ | 10.386 × $10^{-18}$J | 7.235 × $10^{-18}$J | 4.539 × $10^{-18}$J |
| $E_{C3}$ | $4J_{\alpha\beta} + 4J_{\alpha\beta2} + 2J_{\beta\beta} + J_{\beta\beta2}$ | 10.432 × $10^{-18}$J | 7.273 × $10^{-18}$J | 4.571 × $10^{-18}$J |
| $E_{C4}$ | $4J_{\alpha\beta} + 4J_{\alpha\beta2} + 4J_{\beta\beta} + 2J_{\beta\beta2}$ | 10.662 × $10^{-18}$J | 7.463 × $10^{-18}$J | 4.731 × $10^{-18}$J |

TABLE. S1. Energies of nearest-neighbor and next-nearest-neighbor couplings. The energies are calculated by using the micromagnetic simulation program MuMax3. The couplings are defined in Fig. S6a.